\documentclass[aps,amsmath,amssymb,showkeys,showpacs]{revtex4}
\usepackage[dvips]{graphicx,color}
\usepackage{times}
\usepackage{mathrsfs}
\usepackage{amsmath}
\usepackage{graphicx}
\usepackage{epsfig}
\usepackage{dcolumn}
\usepackage{bm}
\begin{document}
\title{Cosmological dynamics of $f(R)$ gravity scalar degree of freedom in Einstein frame}
\author{ Umananda Dev Goswami\footnote{umananda2@gmail.com} and Kabita Deka}
\affiliation{Department of Physics, Dibrugarh University,
Dibrugarh 786004, Assam, India}
\begin{abstract}
$f(R)$ gravity models belong to an important class of modified gravity
models where the late time cosmic accelerated expansion is considered as the
manifestation of the large scale modification of the force of gravity. $f(R)$
gravity models can be expressed in terms of a scalar degree of freedom by
redefinition of model's variable. The conformal transformation of the action
from Jordan frame to Einstein frame makes the scalar degree of freedom more
explicit and can be studied conveniently. We have investigated the features of
the scalar degree of freedoms and the consequent cosmological implications
of the power-law ($\xi R^n$) and the Starobinsky (disappearing cosmological 
constant) $f(R)$ gravity models numerically in the Einstein frame. Both the 
models show interesting behaviour of their scalar degree of freedom and could 
produce the accelerated expansion of the Universe in the Einstein frame with 
the negative equation of state of the scalar field. However, the scalar field 
potential for the power-law model is the well behaved function of the field, 
whereas the potential becomes flat for higher value of field in the case of 
the Starobinsky model. Moreover, the equation of state of the scalar field for 
the power-law model is always negative and less than $-1/3$, which corresponds 
to the behaviour of the dark energy, that produces the accelerated expansion of 
the Universe. This is not always the case for the Starobinsky model. At late 
times, the Starobinsky model behaves as cosmological constant $\Lambda$ as 
behaves by power-law model for the values of $n\rightarrow 2$ at all times.
\end{abstract}


\maketitle
\section{Introduction}
The discovery of the late time accelerated expansion of the Universe 
\cite{Riess1, Riess2, Riess3} demands a new theory as the General Theory of 
Relativity (GTR) could not provide any explanation to this phenomenon within 
its ambit. However, because of all the successful test performed so far on GTR 
for the low scales, attempts have been made to modify GTR in order to account 
the late time behaviour of the cosmic expansion. There are two main conceptual 
approaches that have been led to the modification of GTR. In one approach, a 
new scalar degree of freedom is incorporated in the energy-momentum tensor on 
the right hand side of the Einstein's equation, which is dubbed as dark energy 
for its exotic behaviour with large negative pressure \cite{review11,review12,
review13,review14,review15,review16,review17,review2,review31,review32,
review33,review34}. Initially cosmological constant was considered as the 
source of dark energy which was faced with fine tuning problem of an 
acceptable level. In recent years, as an alternative to the cosmological 
constant, a variety of scalar field models is proposed to provide a viable 
explanation for the phenomenon of late time cosmic acceleration \cite{Zlatev1,
Zlatev2,Zlatev3,Armendariz1,Armendariz2,Armendariz3,Kamenshchik1,Kamenshchik2,
Kamenshchik3,Padmanabhan1,Padmanabhan2,Garousi1,Garousi2}. In the other 
approach, the left hand side of the Einstein's equation can be modified by 
considering the late time cosmic accelerated expansion is due to the large 
scale modification of the force of gravity, which leads to the modified 
gravity theories \cite{Review01,Review02,Review03,Review}. 
It needs to be mentioned that, there are many other alternative attempts to
understand the physical mechanism of this late time behavior of the universe
(e.g. see \cite{sen1,sen2,sen3,sen4,sen5,sen6,sen7,sen8,sen9,sen10,sen11}). 

The simplest class of modified gravity theories is the $f(R)$ gravity theories 
in which Einstein gravity is modified by replacing the Ricci curvature scalar 
$R$ by an arbitrary curvature function $f(R)$. In the special case when $f(R) 
\rightarrow R$, $f(R)$ gravity converge to GTR without a cosmological 
constant \cite{Review01,Review02,Review03,Review}. In last few years, the 
$f(R)$ gravity models have been studied extensively in the cosmological and 
astrophysical aspects because of a number of very interesting results 
(see \cite{Abha} and see references there in). There are many $f(R)$ gravity 
models which could produce the late time cosmic acceleration, but all of them 
are not cosmologically viable and many of them also suffered from the 
singularity and stability problem \cite{Andrei,Amendola,Odin1,Odin2}. So 
certain restrictions have to be imposed on $f(R)$ gravity models to be 
linearly stable and cosmologically viable. Some of such models can be found in 
\cite{Hu1,Hu2,Hu3,Miranda,Odintsov1,Odintsov2}.

It is interesting that a new scalar degree of freedom (sometimes referred as 
{\it scalaron}) appears in $f(R)$ theories of gravity due to redefinition of 
model's variable. The conformal transformation of the metric to the Einstein 
frame makes it explicit in the action \cite{Whitt, Maeda, Wands, Chiba, 
Miranda,Andrei,Abha,Review,Nojiri}. So it is possible to extend $f(R)$ gravity 
to generalized Brans-Dicke (BD) theory with a field potential and an arbitrary 
BD parameter. In $f(R)$ gravity, the scalar field is coupled to 
non-relativistic matter (dark matter, baryons) with a universal coupling 
constant $-1/\sqrt{6}$, and hence as far as the scalar field is concerned it 
is convenient work on $f(R)$ gravity in the Einstein frame, in which a 
canonical scalar field is coupled to non-relativistic matter. However, 
conventionally, the Jordan frame, where the baryons obey the standard 
continuity equation ($\rho_m \propto a^{-3})$, is considered as the physical 
frame in which physical quantities are compared with observations and 
experiments. We are treating essentially the same physics in both frames but 
with different time and length scales that give rise to the apparent 
difference between the observables in two frames. This apparent difference can 
be overlooked by going back to the Jordan frame when we work on the Einstein 
frame for some convenience \cite{Review,Reviewad1,Reviewad2}.

In this work we have studied the $\xi R^n$ and the Starobinsky disappearing
cosmological constant $f(R)$ gravity \cite{Starobinsky} models using their 
scalar degree of freedoms in the Einstein frame in the 
Friedmann-Lema\^itre-Robertson-Walker (FLRW) background. The basic aim
of this study is to see numerically the features of the scalar field coupled
to these $f(R)$ gravity models and hence to see the consequent evolution of the
cosmological parameters in the Einstein frame. We have considered these two
models because: (i) The $f(R) = \xi R^n$ model has exact power-law solutions 
and as a result the singularity is not appear in it as suffered by other 
$f(R)$ models. This is manifested in the scalar degree of freedom as the scalar 
field potential of this model is well behaved function of the field 
\cite{Goheer}. (ii) The Starobinsky $f(R)$ gravity model with disappearing 
cosmological constant \cite{Starobinsky} is a viable cosmological model within
the framework of modified gravity which has been studied extensively in last
few years in different cosmological and astrophysical context 
\cite{Abha,Andrei,Lee1,Lee2,Lee3,Lee4}. This will help us to 
compare the results of $\xi R^n$ gravity model with results of the standard
model \cite{Starobinsky}. We organized this work as follows: In Sec.II we 
discuss the basics of formalism of the $f(R)$ gravity theory in the Jordan 
frame and then we transform the formalism to the Einstein frame using the 
conformal transformations to developed field equations of the scalar degree 
of freedom. The numerical analysis of the work based on the formalism 
developed in the Sec.II is done in the Sec.III for our models as mentioned 
above. Finally in the Sec.IV we concludes the results of our analysis.
                 
\section{$f(R)$ gravity formalism in Einstein frame}
In this section we will develop the field equations of $f(R)$ gravity in 
Einstein frame starting from the action of $f(R)$ gravity in the Jordan frame. 
The $f(R)$ gravity action in the Jordan frame is given by 
\cite{Review01,Review02,Review03,Review},
\begin{equation}
S = \frac{1}{2\kappa^2}\int d^4x\sqrt{-g}f(R) + 
\int d^4\mathcal{L}_m(g_{\mu\nu}, \Psi_m),
\label{eq1}
\end{equation}
where $\kappa^2 = 8\pi G$, $g$ is the determinant of the metric $g_{\mu\nu}$, 
and $\mathcal{L}_m$ is the matter Lagrangian, which is the function of the
metric $g_{\mu\nu}$ and the matter fields $\Psi_m$. The variation of the 
action (\ref{eq1}) with respect to $g_{\mu\nu}$ the leads to the equation of 
motion \cite{Review, Miranda, Andrei}:
\begin{equation}
f^{\prime}R_{\mu\nu} -\nabla_{\mu\nu}f^{\prime} +\left(\Box f^\prime - 
\frac{1}{2}f\right)g_{\mu\nu} = \kappa^2 T_{\mu\nu},
\label{eq2}
\end{equation}
where $(\prime)$ denotes the derivatives with respect to $R$. The term 
$\Box f^\prime$ does not vanish in modified gravity, which means that 
there is a propagating scalar degree of freedom, defined as $\varphi
 = f^\prime$, whose dynamics is governed by the trace of the equation 
(\ref{eq2}) given by,
\begin{equation}
3\Box f^\prime + f^\prime R - 2f = \kappa^2T.
\label{eq3}
\end{equation}      
In practice, it is usually difficult to invert the definition of the scalar
degree of freedom explicitly for a given $f(R)$ and only convenient to
determine effective potential in a parametric form \cite{Andrei}. However, 
using the conformal transformation to the Einstein frame it is possible to 
express the scalar degree of freedom of a given $f(R)$ in the explicit 
form \cite{Andrei}. In view this
we now switch to the Einstein frame under the following conformal 
transformation \cite{Review, Barrow1,Barrow2}:
\begin{equation}
\tilde{g}_{\mu\nu} = \Omega^2g_{\mu\nu},
\label{eq4}
\end{equation}
where $\Omega^2$ is the conformal factor and the quantities in the Einstein 
frame are represented by a tilde over them. In these two frames the Ricci 
scalars have the following relation:
\begin{equation}
R = \Omega^2(\tilde{R} + 6\tilde{\Box}\omega - 
6\tilde{g}^{\mu\nu}\partial_\mu\omega\partial_\nu\omega),
\label{eq5}
\end{equation}
where
\begin{equation}
\omega \equiv ln\Omega,\;\;\; \partial_\mu\omega \equiv \frac{\partial}{\partial\tilde{x}^\mu},\;\;\; \tilde{\Box}\omega \equiv \frac{1}{\sqrt{-g}}\partial(\sqrt{-\tilde{g}}\tilde{g}^{\mu\nu}\partial_{\nu}\omega),
\label{eq6}
\end{equation}

For convenient we rewrite the action (\ref{eq1}) in the form \cite{Review}:
\begin{equation}
S = \int d^4x\sqrt{-g}\left(\frac{1}{2\kappa^2}f^\prime R - U\right) +
\int d^4x\mathcal{L}_m(g_{\mu\nu}, \Psi_m),
\label{eq7}
\end{equation}
where
\begin{equation}
U = \frac{f^\prime R -f}{2\kappa^2}.
\label{eq8}
\end{equation}
Using Eq. (\ref{eq5}) and the relation 
$\sqrt{-g} = \Omega^{-4}\sqrt{-\tilde{g}}$, the action (\ref{eq7}) can be 
written as \cite{Review}
\begin{equation}
S = \int d^4x\sqrt{-\tilde{g}}\left[\frac{1}{2\kappa^2}f^\prime\Omega^{-2}(\tilde{R} + 6\tilde{\Box}\omega - 6\tilde{g}^{\mu\nu}\partial_{\mu}\omega\partial_\nu\omega)-\Omega^{-4}U \right] + \int d^4x\mathcal{L}_m(\Omega^{-2}\tilde{g}_{\mu\nu}, \Psi_m).
\label{eq9}
\end{equation}
The integral $\int d^4x\sqrt{-\tilde{g}}\tilde{\Box}\omega$ can be made to 
vanish on account of the Gauss's theorem by using equation (\ref{eq6}). Now
if make a choice \cite{Review,Barrow1,Barrow2},
\begin{equation}
\Omega^2 = f^\prime,
\label{eq10}
\end{equation}
for $f^\prime > 0$ and introduce a new scalar field $\phi$ defined by
\begin{equation}
\kappa\phi \equiv \sqrt{3/2}\;\mbox{ln}\; f^\prime,
\label{eq11}
\end{equation}
then the action in the Einstein frame can be expressed as \cite{Review}
\begin{equation}
S_E = \int d^4x\sqrt{-\tilde{g}}\left[\frac{1}{2\kappa^2}\tilde{R} - \frac{1}{2}\tilde{g}^{\mu\nu}\partial_\mu\phi\partial_\nu\phi -V(\phi)\right] + 
\int d^4x\mathcal{L}_m(F^{-1}(\phi)\tilde{g}_{\mu\nu}, \Psi_m),
\label{eq12}
\end{equation}
where
\begin{equation}
V(\phi) = \frac{U}{f'^{2}} = \frac{f^\prime R - f}{2\kappa^2f'^2}
\label{eq13}
\end{equation}
is the potential of the field $\phi$. Here the conformal factor is
\begin{equation}
\Omega^2 = f^\prime = exp(\sqrt{2/3}\kappa\phi),
\label{eq14}
\end{equation}
which is field dependent. The Lagrangian density of the field $\phi$ is given
by
\begin{equation}
\mathcal{L}_\phi = -\frac{1}{2}\tilde{g}^{\mu\nu}\partial_\mu\phi\partial_\nu\phi - V(\phi).
\label{eq15}
\end{equation}
From the matter action (\ref{eq12}) it is clear that the scalar field $\phi$ 
is directly coupled to matter in the Einstein frame, which can be seen more
explicitly if we take the variation of the action (\ref{eq12}) with respect to
the field $\phi$ \cite{Review}. However, we are not interested in this 
context, instead we ignore the matter field to derive the scalar field 
equations without coupling to the matter. Thus taking the variation of the 
action (\ref{eq12}) with respect to the field $\phi$, we obtain the field 
equations without the presence of matter as \cite{Review}
\begin{equation}
\ddot{\phi} + 3\tilde{H}\dot{\phi} + V_{,\phi} = 0,
\label{eq16}
\end{equation}
\begin{equation}
\tilde{H}^2 = \frac{\kappa^2}{3}\left[\frac{1}{2}\dot{\phi}^2 + V(\phi)\right].
\label{eq17}
\end{equation}
Here the dots over $\phi$ indicate the derivatives with respect to conformal
cosmic time (i.e. cosmic time in Einstein frame) $\tilde{t}$. $\tilde{H} = 
\frac{\tilde{\dot{a}}}{\tilde{a}}$ is the Hubble parameter in the Einstein 
frame, $\tilde{a}(t)$ being the scale factor in the same frame. If we consider 
the spatially flat Friedmann-Lema\^{i}tre-Robertson-Walker (FLRW) spacetime 
with a comic time-dependent scale factor $a(t)$ and a metric, viz.,
\begin{equation}
ds^2 = g_{\mu\nu}dx^\mu dx^\nu = -dt^2 + a^2(t)d{\bf x}^2
\label{eq18}
\end{equation}
in the Jordan frame, then the metric in the Einstein frame may be given as
\begin{equation}
d\tilde{s}^2 = \Omega^2ds^2 = f^\prime(-dt^2 + a^2(t)d{\bf x}^2) = 
-d\tilde{t}^2 + \tilde{a}^2(\tilde{t})d{\bf x}^2,
\label{eq19}
\end{equation}
which leads to the relations of the cosmic time and scale factor in two frame
as
\begin{equation}
d\tilde{t} = \sqrt{f^\prime}dt,\;\;\; \tilde{a} = \sqrt{f^\prime}a,\;\; 
\mbox{for}\;\; \sqrt{f^\prime}> 0.
\label{eq20}
\end{equation}
Finally we define the energy density and the pressure of the scalar field 
as \cite{Review}
\begin{equation}
\tilde{\rho}_\phi = \frac{\dot{\phi}^2}{2} + V(\phi),
\label{eq21}
\end{equation}
\begin{equation}
\tilde{P}_\phi = \frac{\dot{\phi}^2}{2} - V(\phi).
\label{eq22}
\end{equation}  
It is evident from the above formalism that, in 
Einstein frame the scalar degree of freedom for a given $f(R)$ model can be
studied explicitly and conveniently. In the next section we will apply the 
results of this formalism to study our models of interest.

\section{Scalar field dynamics of $f(R)$ gravity model}
Using the formalism of the previous section, in this section we will study 
numerically different features of the scalar field and consequent cosmological 
implications of the power-law  and the Starobinsky $f(R)$ gravity models as
follows:

\subsection{Power-law $f(R)$ gravity model}
The general power-law $f(R)$ gravity model is given by 
\begin{equation}
f(R) = \xi R^n,
\label{eq23}
\end{equation}  
where $\xi$ and $n$ are model parameters. We use this $f(R)$
gravity model in this study because of its unique character for the existence 
of power-law solutions and hence absence of the singularity problem as 
mentioned in the Sec.I \cite{Goheer}. Here our main emphasis is to see the 
explicit behavior of scalar degree of freedom possessed by the $f(R)$ gravity 
model and its usefulness to study the cosmological evolution.

Using the equation (\ref{eq11}), the scalar field $\phi$ corresponding to the
model (\ref{eq23}) can be expressed as
\begin{equation}
\phi = \sqrt{\frac{3}{2}}\frac{1}{\kappa}lnf^\prime = \sqrt{\frac{3}{2}}\frac{1}{\kappa}ln(\xi nR^{n-1}).
\label{eq24}
\end{equation}
It is clear from this expression that, for the positive valued of $\xi$ and 
$n\; (> 1)$, the scalar field $\phi$ is the increasing function of the Ricci 
curvature scalar $R$ and hence in this case, when $R\rightarrow \infty$, 
$\phi\rightarrow \infty$. This indicates that, under this condition there exist
no singularity in the scalar field for any finite value of the curvature 
scalar $R$.
Using this expression for $\phi$, the field potential (\ref{eq13}) takes the
form:
\begin{equation}
V(\phi) = \frac{n-1}{2\kappa^2\xi n^2}\left(\frac{exp(\sqrt{2/3}\kappa\phi)}{\xi n}\right)^{\frac{2-n}{n-1}}.
\label{eq25}
\end{equation}
This equation for the field potential shows that the value of $n$ should be 
within two possible ranges: (i) $n<1$ and (ii) $1<n<2$ for a viable field 
potential. For our numerical simulation in this work we consider the second
range of the values of $n$ and the values of $\xi<1$, in view of the above 
discussion for the scalar field $\phi$. We consider $\kappa =1$
through out this work.  
\begin{figure}[hbt]
\centerline
\centerline{
\includegraphics[width = 7cm, height = 6cm]{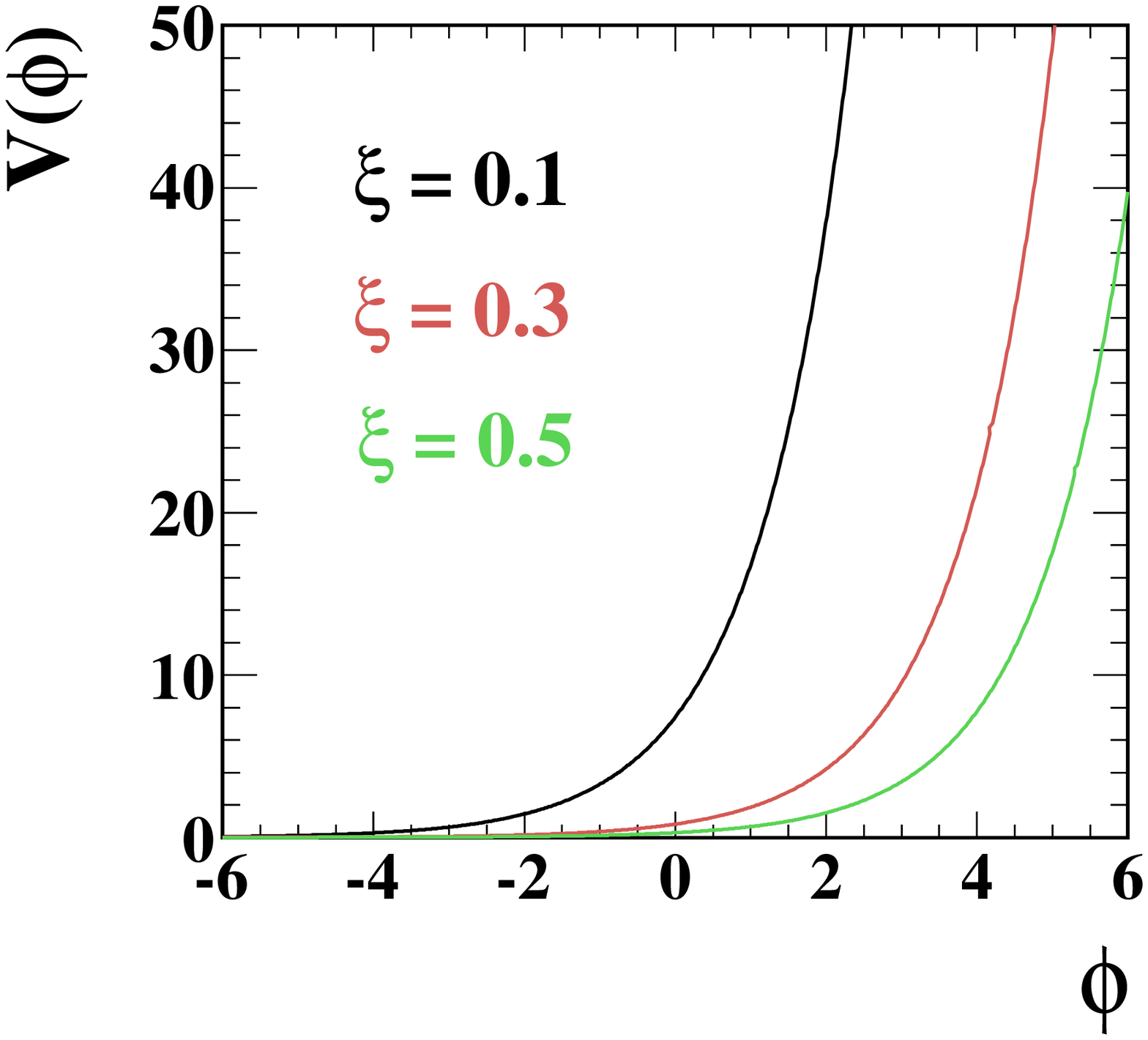}
\includegraphics[width = 7cm, height = 6cm]{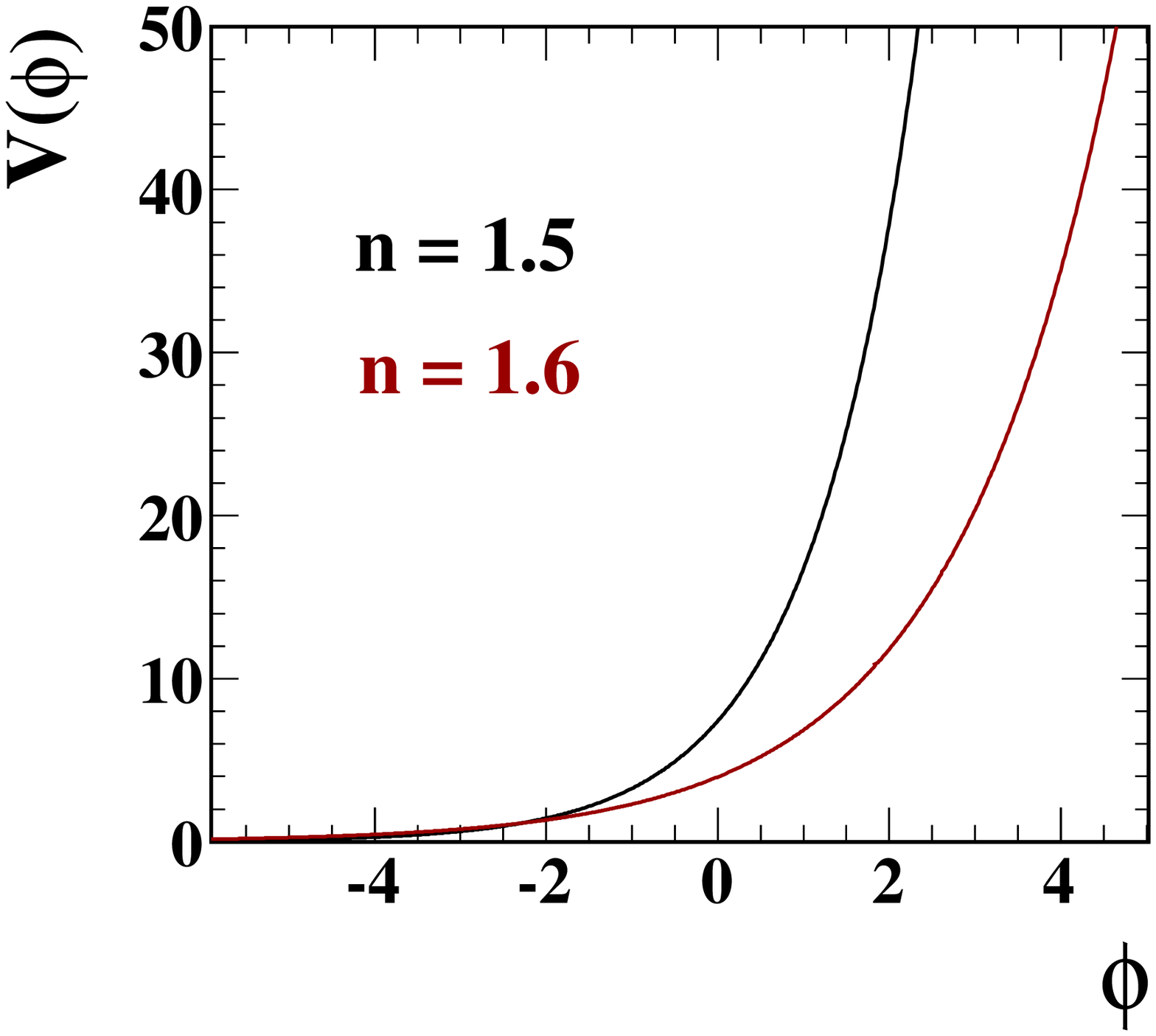}
}
\caption{The variation of field potential
with respect to the field $\phi$. The first panel is for $n = 1.5$ with $\xi =  
0.1$, $0.3$ and $0.5$, whereas the second panel is for $n = 1.5$ and
$1.6$ with $\xi = 0.1$. The value of $\kappa$ is taken as one.}
\label{fig1}
\end{figure}

Fig.\ref{fig1} shows the variation of field potential 
with respect to the field $\phi$. The first panel is for $n = 1.5$ with $\xi = 
0.1$, $0.3$ and $0.5$, whereas the second panel is the plot for $n = 1.5$ and
$1.6$ with $\xi = 0.1$. From this figure it is seen that the potential is 
well behaved function of the field and the potential is small for negative 
value of the field and it increases very fast as soon as the 
field $\phi$ become positive. This process is slowing down when the value of
$\xi$ increases, which is more effective for lower value range of $\xi$, as 
clear from the left panel. On the other hand as seen from
the second panel that there is no significant difference of the potentials for
the different values of $n$ within the range of negative field, but the 
difference becomes more prominent as soon as the field $\phi$ acquired the 
positive values. The effect of $n$ on the value of the potential is similar to 
that for $\xi$, but the effect on higher positive value of $\phi$ is more 
noticeable in this case than the case for $\xi$.  
\begin{figure}[hbt]
\centerline
\centerline{
\includegraphics[width = 5.5cm, height = 5cm]{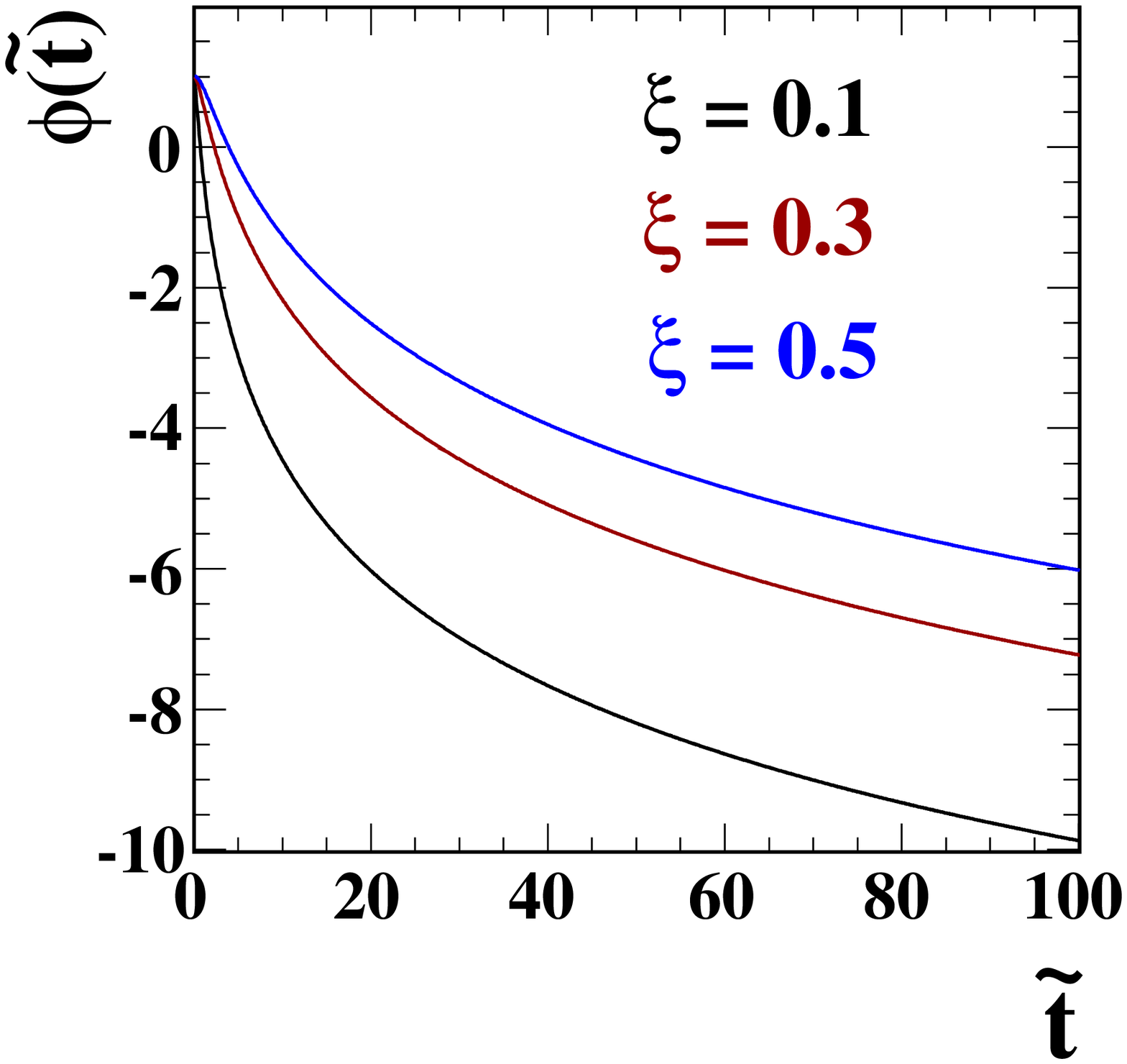}
\includegraphics[width = 5.5cm, height = 5cm]{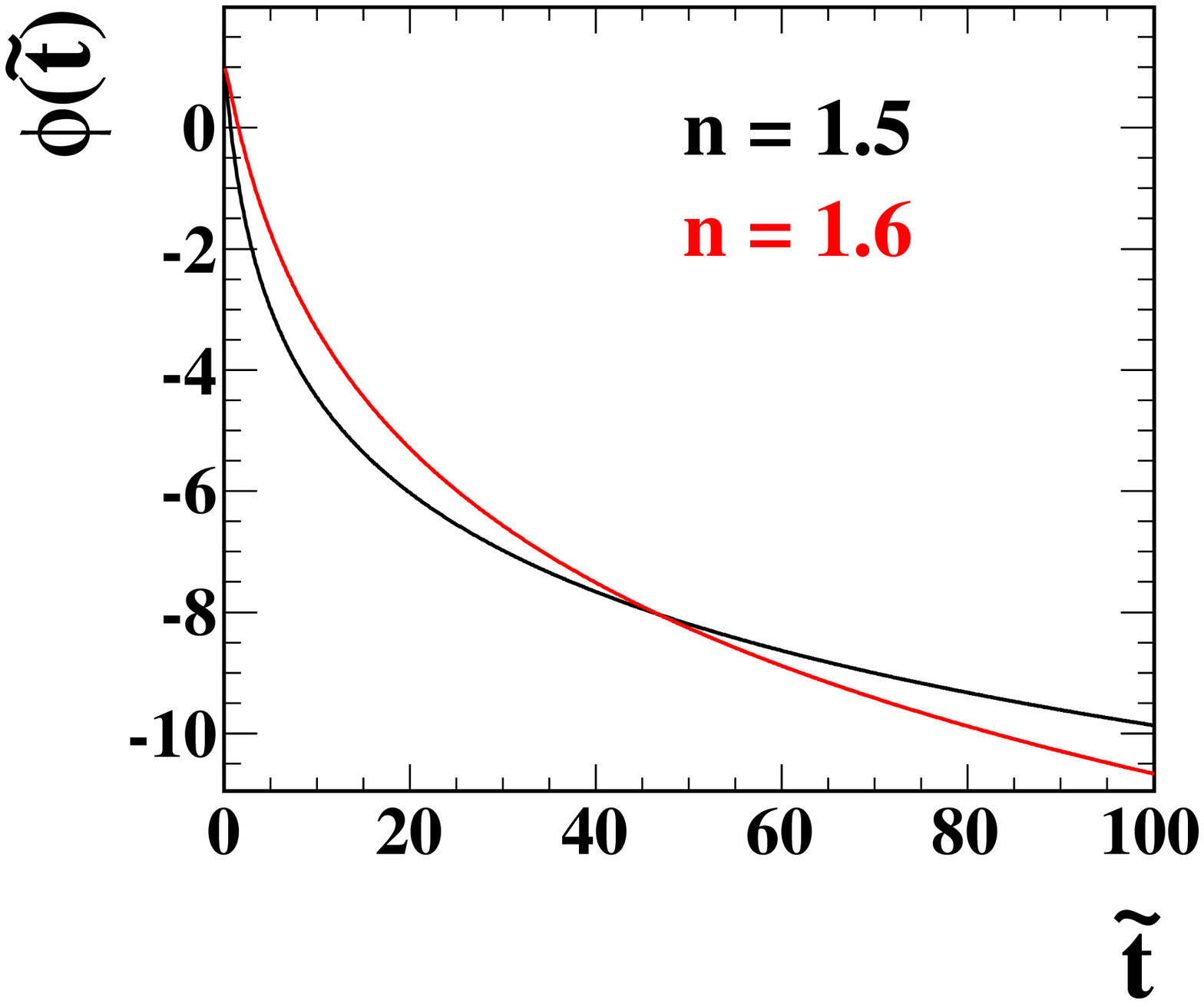}
\includegraphics[width = 5.5cm, height = 4.905cm]{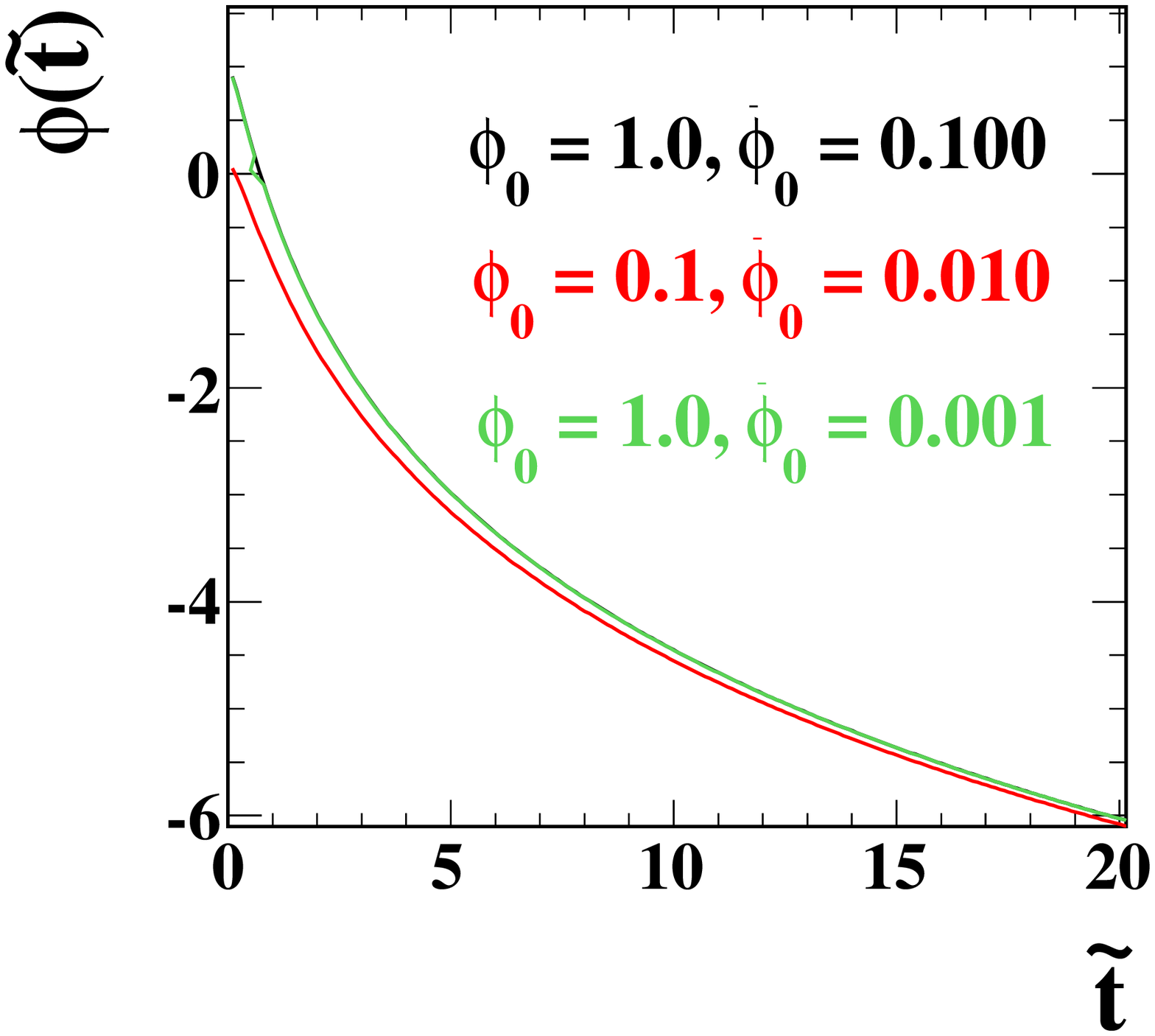}
}
\caption{Numerical solutions of equations (\ref{eq16}) and (\ref{eq17}) using
the field potential (\ref{eq25}) for different values of $n$ and $\xi$. The
first panel is for $n=1.5$ with different values of $\xi$. The second panel
is for $\xi = 0.1$ with two values of $n$. In both these panels $\phi_0=1.0$ and
$\dot{\phi}_0 = 0.1$ are used. While the last panel is for different initial
values of $\phi$ and $\dot{\phi}$ with $n = 1.5$ and $\xi = 0.1$. The value of
$\kappa$ is taken as one.}
\label{fig2}
\end{figure}                    

Using the potential (\ref{eq25}), the coupled field equations (\ref{eq16}) and
(\ref{eq17}) are solved numerically for different values of $n$ and $\xi$ as 
mentioned
above. The results of the numerical solution are shown in the Fig.\ref{fig2}.
The first panel shows the numerical solutions for $n= 1.5$ with $\xi = 0.1, 0.2$
and $0.3$. The second panel is for $n= 1.5$ and $1.6$ with 
$\xi = 0.1$. In both cases the initial field value and the initial field 
velocity are taken as $1.0$ and $0.1$ respectively. As clear from the first 
panel that the value of the field $\phi$ falls towards to the negative values
starting from its initial value as time increases. For the initial period of
cosmic time $\phi$ falls very fast and this tendency slows down as time passage.
For higher values of $\xi$ the overall tendency of the $\phi$ to fall towards
the negative values is less in comparison to its lower values. This is similar
to the nature of the field potential with respect to $\xi$. Form the second
panel it is observed that the field falls
faster initially for the smaller value of $n$ as compared to the higher value,
but the situation reversed in the latter cosmic times. During the time before 
the situation is reversed, the difference of the field $\phi$ is not very high, 
however after this period the difference become increasingly significant as 
time elapsed. We have also studied the effect of the initial field and field
velocity on the time evolution of the field, whose results are shown in the
last panel of the Fig.\ref{fig2}. This plot is for $n = 1.5$ and $\xi = 0.1$
with $\phi_0 = 1.0, 0.1$ and $\dot{\phi}_0 = 0.1, 0.01, 0.001$. We have seen 
that the initial values of $\phi$ and $\dot{\phi}$ does not have much 
noticeable effect on the time evolution of $\phi$. Initial value of  $\phi$ 
has slight
effect during initial period, which eliminates gradually in latter times. On 
the other hand the initial values of $\dot{\phi}$ does not have any effect
on the time evolution of $\phi$.
\begin{figure}[hbt]
\centerline
\centerline{
\includegraphics[width = 5.5cm, height = 5cm]{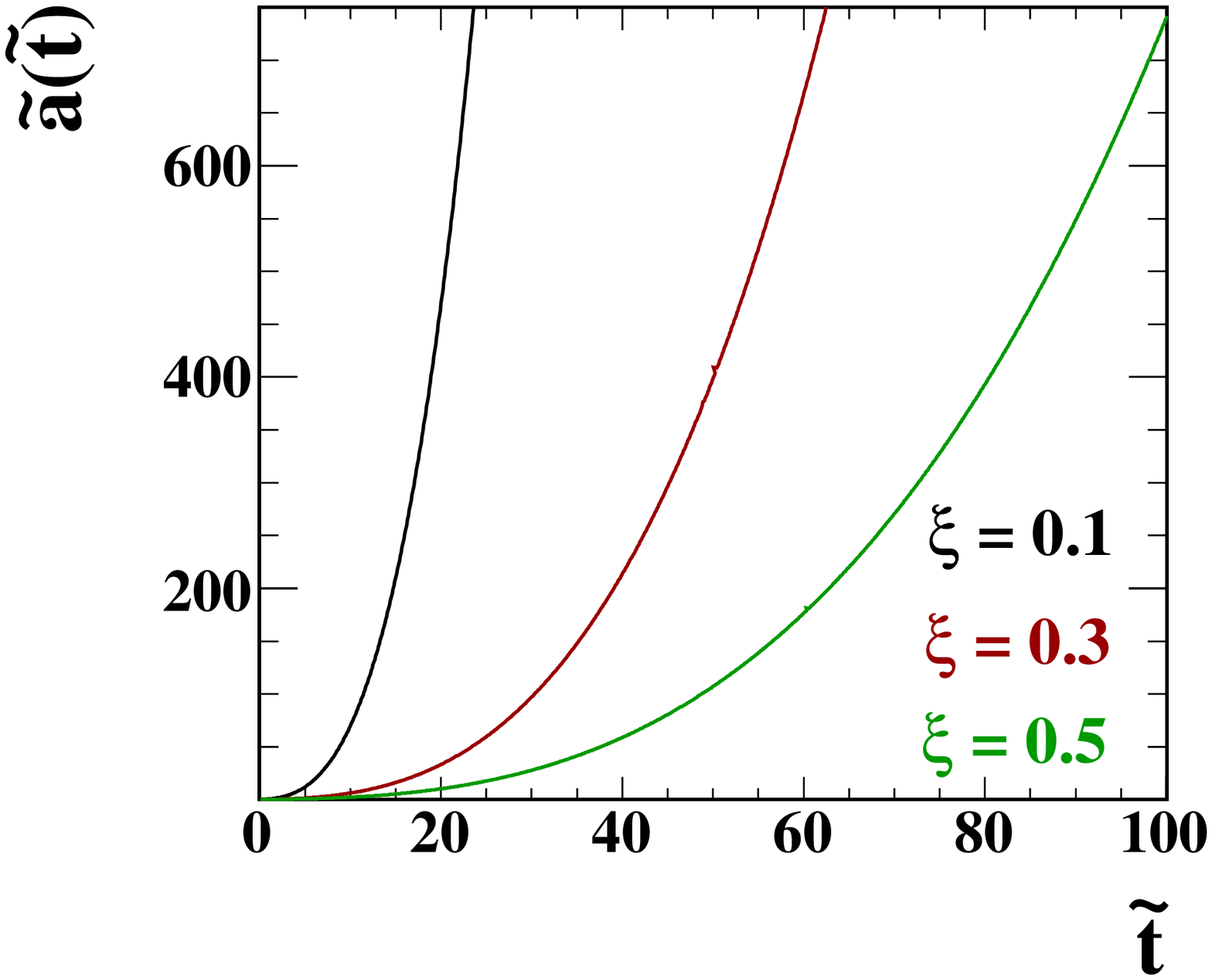}
\includegraphics[width = 5.5cm, height = 5cm]{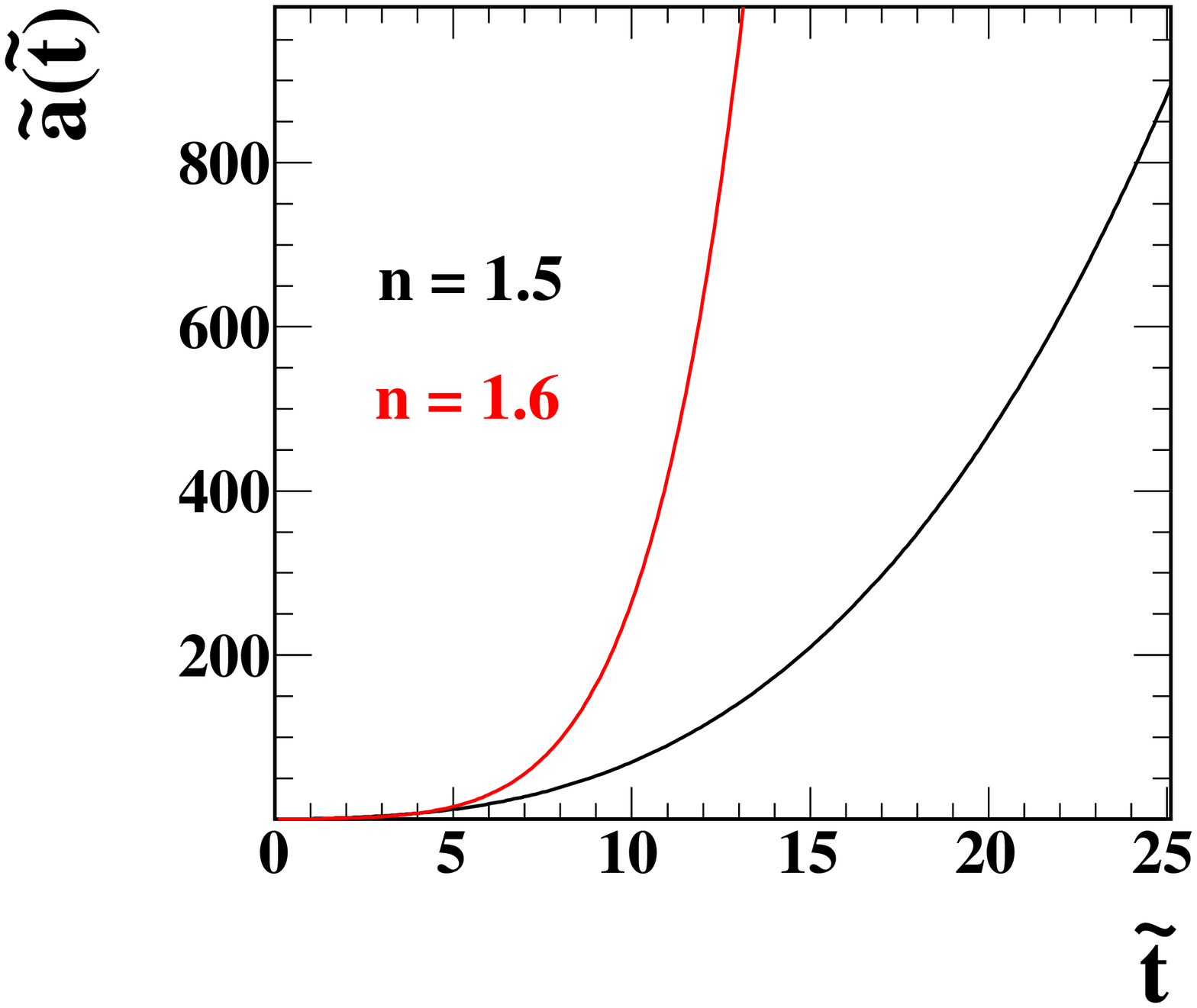}
\includegraphics[width = 5.5cm, height = 5cm]{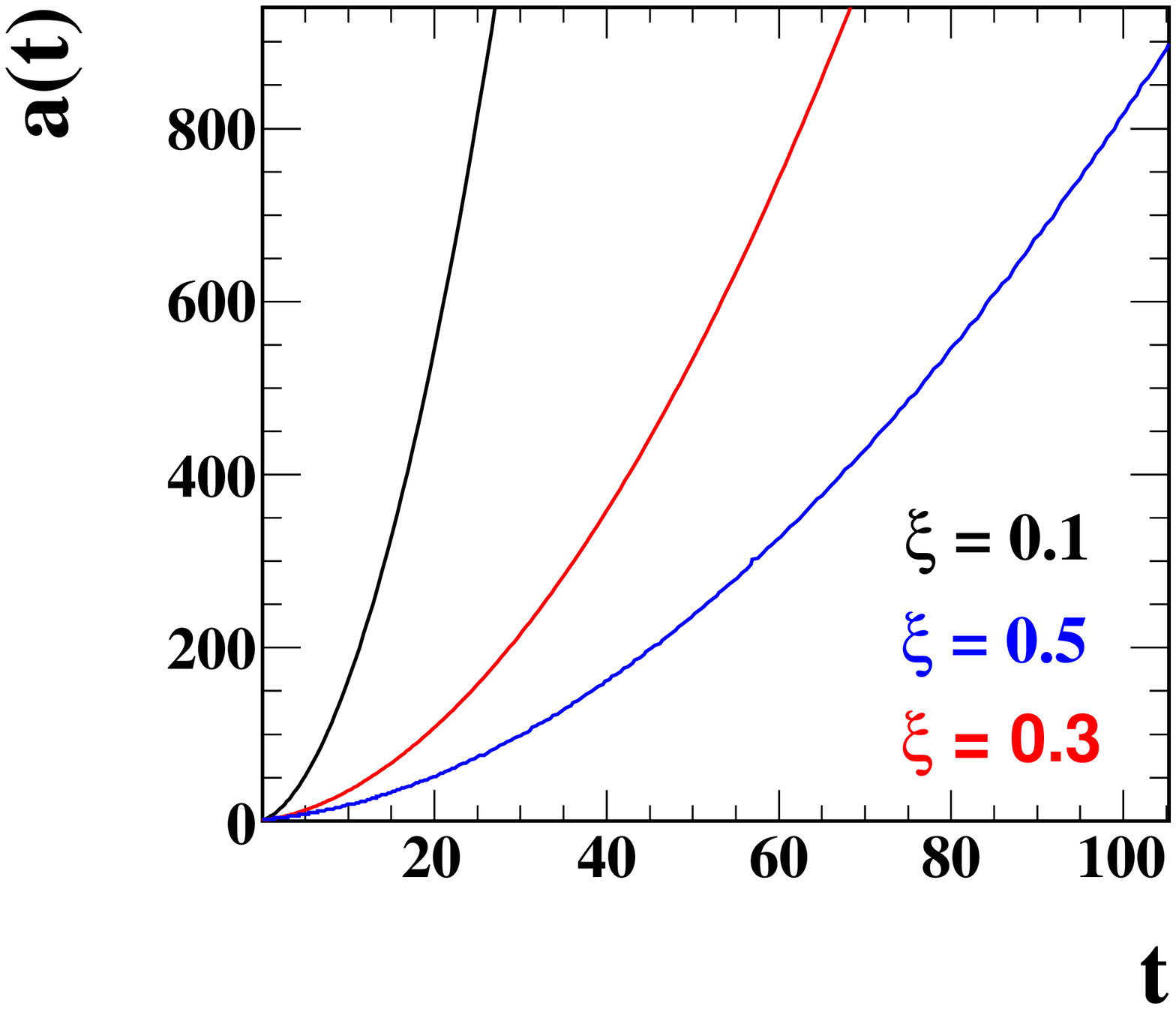}
}
\caption{First two panels: Evolution of the scale factor in the Einstein frame
obtained from the numerical solutions of equations (\ref{eq16}) and
(\ref{eq17}) using the field potential (\ref{eq25}) for different values of
$n$ and $\xi$. The first panel is for $n=1.5$ with different values of $\xi$.
Whereas the second panel is for $\xi = 0.1$ with two vales of $n$. In both
these panel $\phi_0=1.0$ and $\dot{\phi}_0 = 0.1$ are used. The value of
$\kappa$ is taken as one. Last panel: Evolution of the scale factor in Jordan
frame obtained from the first panel by using the equation (\ref{eq20}).}
\label{fig3}
\end{figure}

From these numerical solutions we have calculated the scale factors 
$\tilde{a}(\tilde{t})$ in Einstein frame for different values of $n$ and $\xi$ 
as mentioned above. The (cosmic) time evolution of the scale factors 
obtained from these calculation are shown in the the first two panels of the 
Fig.\ref{fig3}. The first panel of this figure shows the time evolution of 
the scale factor for $n=1.5$ with different values of $\xi$. Whereas the second
one is for $\xi =0.1$ with two different values of $n$. The initial parameters 
are same as for the first two panels of the  Fig.\ref{fig2}. It is observed 
from these two panels of this figure that the $f(R) = \xi R^n$ gravity model 
leads the very fast expansion of the Universe in the Einstein frame similar to 
the exponential expansion. The rate of expansion of the Universe slowed down 
gradually for higher values of $\xi$ in comparison to its lower 
values. During initial period this process is slow, but become more significant as time passage. On the other hand the expansion is slower for lower value of 
$n$ in comparison to the case for the higher value of $n$ as clear from the 
second panel. In the initial period the difference of the values of the scale 
factor is very lass, but becomes more prominent in the latter times. 

To see explicitly, whether the acceleration of the Universe always occurs 
simultaneously in the the Jordan frame, as in the case of the Einstein frame,
we have calculated the cosmic time $t$ and the scale factor $a(t)$ in the 
Jordan frame using the equation (\ref{eq20}) for all three conditions of the
first panel of the Fig.\ref{fig3}, which is shown in the last panel of the 
same figure. This panel shows that, this power law model generates very fast
expansion of the Universe is the Jordan frame also similar to that in the
Einstein frame, but with a slightly different magnitude and in pattern in the 
parameter space. Where it is seen that, although the trend of variation of the
scale factor, with respect to the parameter $\xi$, remains same as in the case 
of the Einstein frame, the initial slowing down of the variation of the scale 
factor reduces considerably in the Jordan frame. Moreover, the magnitude of 
the acceleration of the
Universe is slightly higher in the Jordan frame than that in the Einstein 
frame. Obviously, with the increasing value of $n$, the expansion of the 
Universe in the Jordan frame will follow the pattern in accordance with the 
second plot shown for the Einstein frame.                      
\begin{figure}[hbt]
\centerline
\centerline{
\includegraphics[width = 7cm, height = 6cm]{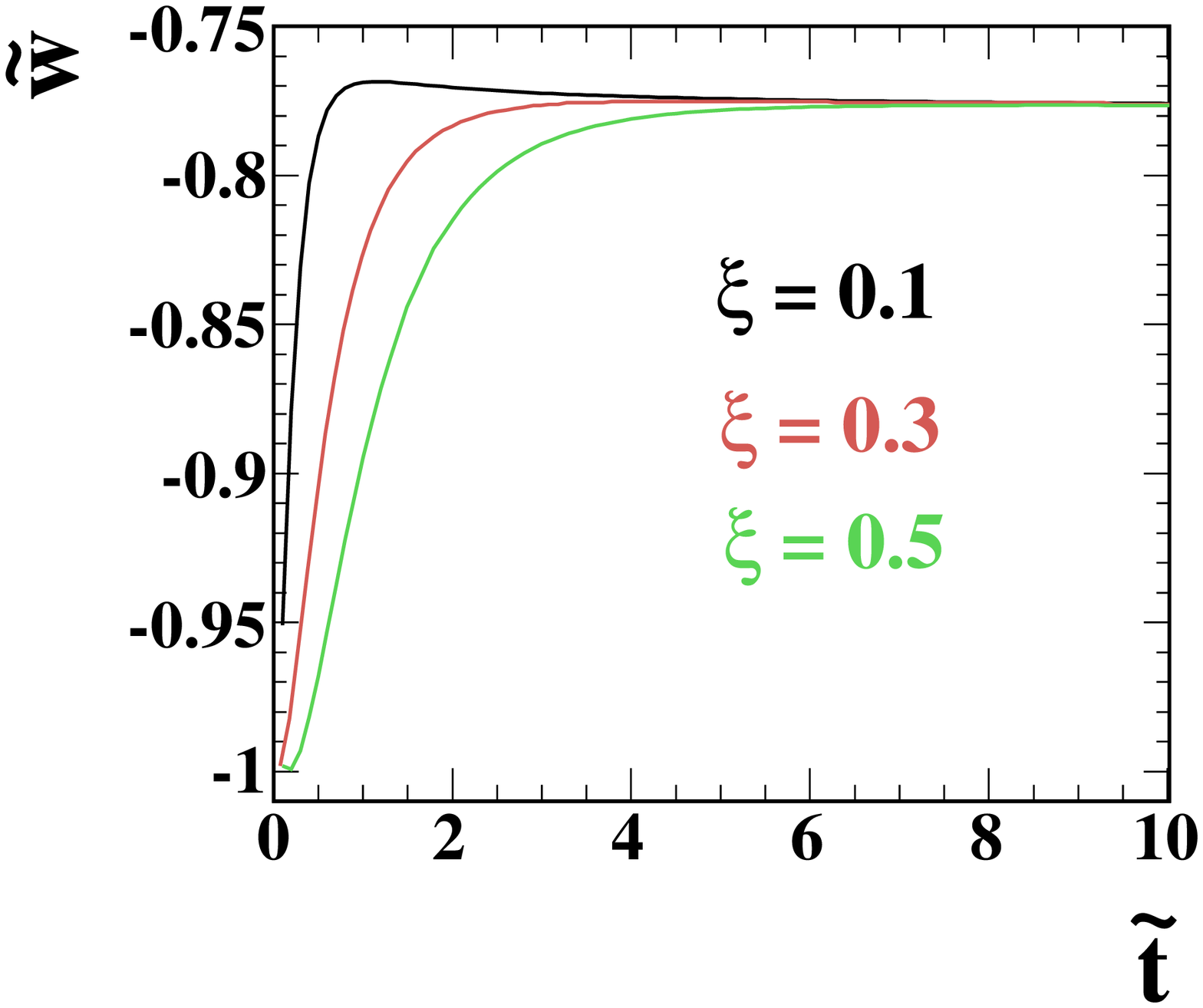}
\includegraphics[width = 7cm, height = 6cm]{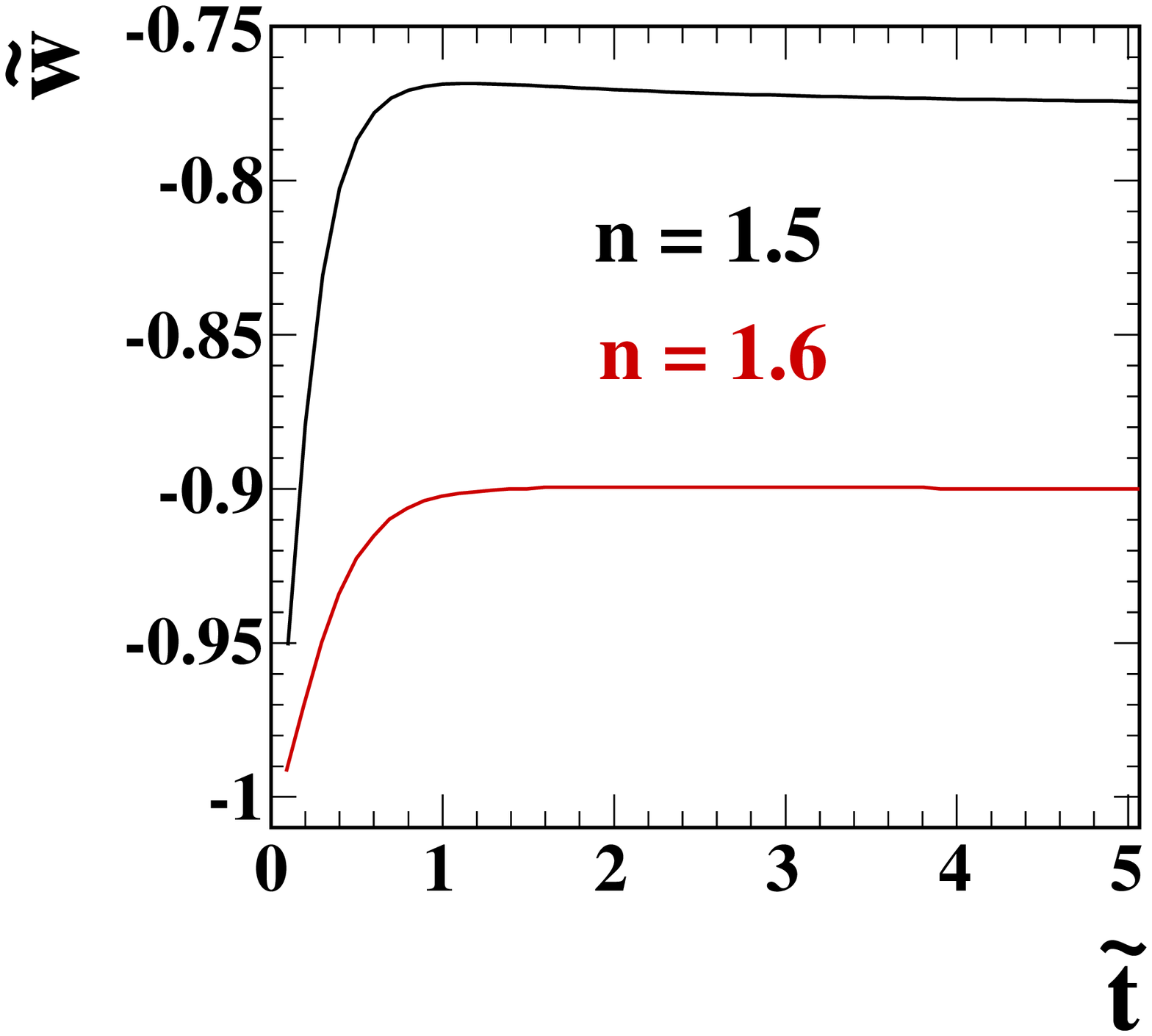}
}
\caption{Equation of state of the scalar field in the Einstein frame obtained
from the
numerical solutions of equations (\ref{eq16}) and (\ref{eq17}) using
the field potential (\ref{eq25}) for different values of $n$ and $\xi$. As in
the previous figure the first panel is for $n=1.5$ with different values of
$\xi$. Whereas the second panel is for $\xi = 0.1$ with two vales of $n$. All
initial conditions and model parameters are same with the previous figure.}
\label{fig4}
\end{figure}
     
At last we have calculated the equation of state $\tilde{w} = 
\tilde{\rho}_{\phi}/\tilde{P}_{\phi}$ in Einstein frame using 
equations (\ref{eq21}) and (\ref{eq22}) from the numerical solutions mentioned
above, which is shown in the Fig.\ref{fig4}. Panels of this figure are for the
same model parameters respectively as for the panels of the Fig.\ref{fig3}. It
is observed from the figure that the equation of state $\tilde{w}$ of the 
scalar 
field $\phi$ for the model (\ref{eq23}) always less than $-0.75$ for all cases 
of our study and for all times. For higher values of $\xi$, $\tilde{w}$ is 
smaller than that for
smaller value of $\xi$ during period for some initial times, but for latter 
times $\tilde{w}$ becomes equal for all $\xi$. But for higher value of $n$, 
$\tilde{w}$ is always smaller than that for the smaller value of $n$ with 
a similar trend with respect to time and $\xi$ as mentioned above. As the 
equation of state is always negative, therefore the model (\ref{eq23}) could 
produced the accelerated expansion of the Universe and it corresponds to 
dark energy, whose equation of state is $w<-1/3$ (in the Jordan frame also the 
equation of state will lay within this range as clear from the last panel of
the Fig.\ref{fig3}). If we increase the 
values of $n$ then the equation of state $\tilde{w}$ in Einstein frame will 
approach to $-1$. This implies that the model (\ref{eq23}) with higher values
of $n$ ($n\rightarrow$2) corresponds to the cosmological constant $\Lambda$.           
\subsection{Starobinsky $f(R)$ gravity model}
The Starobinsky $f(R)$ gravity model with disappearing cosmological constant
\cite{Starobinsky} is,
\begin{equation}
f(R) = R + 
\lambda R_0\left[\left(1 + \frac{R^2}{R_{0}^{2}}\right)^{-n} - 1\right],
\label{eq26}
\end{equation} 
where $n$, $\lambda$  and $R_{0}$ are free parameters. The values of $n$ and 
$\lambda$ are greater than zero, whereas the value of $R_{0}$ is of the
order of the presently observed cosmological constant 
$\Lambda \;(= 8\pi G\rho_{vac})$. This is carefully constructed model to give
viable cosmology and to be satisfied with solar system and laboratory tests.
Notwithstanding, it suffers from the singularity problem \cite{Andrei}, which
could be cured by adding a term $\propto R^2$ to it as observed in 
\cite{Odin1,Odin2} and consequently this observation was followed in 
\cite{Abha}. With
this modification, the Starobinsky model (\ref {eq26}) takes the form:
\begin{equation}
f(R) = R + \frac{\beta}{R_0}R^2 +
\lambda R_0\left[\left(1 + \frac{R^2}{R_{0}^{2}}\right)^{-n} - 1\right],
\label{eq26a}
\end{equation}
where $\beta$ is the positive constant parameter.

The scalar field embodied in the model (\ref{eq26}) which is obtained by using 
the equation (\ref{eq11}) can be written as
\begin{equation}
\phi = \sqrt{\frac{3}{2}}\frac{1}{\kappa}ln f^\prime = 
\sqrt{\frac{3}{2}}\frac{1}{\kappa}ln\left[1 - 2\lambda n\frac{R}{R_0}\left(1 +
\frac{R^2}{R_{0}^2}\right)^{-(n+1)}\right].
\label{eq27}
\end{equation}
From the observation of this equation it is quite clear that, for the given 
values of the parameters $n$ and $\lambda$, the scalar field $\phi$ increases
from the negative side with the increasing value of the Ricci curvature
scalar $R$. So when $R\rightarrow \infty$, $\phi\rightarrow 0$, which is the 
point of singularity. However, if we express the scalar field associated with
the Starobinsky modified model (\ref{eq26a}), it takes the form:
\begin{equation}
\phi = \sqrt{\frac{3}{2}}\frac{1}{\kappa}ln f^\prime = 
\sqrt{\frac{3}{2}}\frac{1}{\kappa}ln\left[1 + \frac{R}{R_0}\left\{2\beta - 
2\lambda n\left(1 + \frac{R^2}{R_{0}^2}\right)^{-(n+1)}\right\}\right].
\label{eq27a}
\end{equation}      
This equation indicates that for the large value of $R$, the second term 
coming from $\propto R^2$ term of the model (\ref{eq26a}) dominates over other 
terms in the expression 
and hence the situation as discussed above for the equation (\ref{eq27}) 
never arise here, but when $R\rightarrow \infty$, $\phi\rightarrow \infty$. 
Thus in this case for a finite curvature, we get finite value of the field
$\phi$. Nevertheless, in the context of our motivation, we will use the 
Strobinsky original model (\ref{eq26}) for the further discussion below. 
We will use the modified model (\ref{eq26a}) only to see the effect on
singularity in the scalar field potential.   

Now, if we use the expression (\ref{eq27}) for $\phi$ in a region where 
$R/R_0 >> 1$, the field potential (\ref{eq13}) for the Starobinsky model 
(\ref{eq26}) becomes,
\begin{equation}
V(\phi) = -\;\lambda \frac{\left[(2n+1)\left(\frac{1\;-\;exp({\sqrt{2/3} \kappa\phi})}{2n\lambda}\right)^{\frac{2n}{2n+1}} - 1\right]}{2\kappa^2 \;exp({\sqrt{8/3}} \kappa\phi)} 
\label{eq28}
\end{equation}
Similarly, using the expression (\ref{eq27a}) for $\phi$ in the same region, 
the field potential (\ref{eq13}) for the modified model (\ref{eq26a}) to a 
good approximation can be written as
\begin{equation}
V(\phi) = \frac{\beta\left(\frac{exp({\sqrt{2/3} \kappa\phi}) \;-\;1}{2\beta}\right)^2 - \lambda\left[\left(\frac{exp({\sqrt{2/3} \kappa\phi}) \;-\;1}{2\beta}\right)^{-2n} -1 \right]}{2\kappa^2\;exp({\sqrt{8/3}} \kappa\phi)}.
\label{eq28a}
\end{equation} 
The expression for the field potential (\ref{eq28})  shows that for the 
validity of the 
Starobinsky model (\ref{eq26}) the values of $n$ and $\lambda$ must be greater
than zero, but may take any values freely beyond it in contrast with the 
power-law model (\ref{eq23}). However, for our numerical work on this model, 
we have taken $n = 1$ and $2$ with the values of $\lambda = 1.2,\;1.5,\; 2.0$ 
and $4.0$. We have followed the same numerical procedures for this model also
that have been used for the case of power-law model.

\begin{figure}[hbt]
\centerline
\centerline{
\includegraphics[width = 5.5cm, height = 5cm]{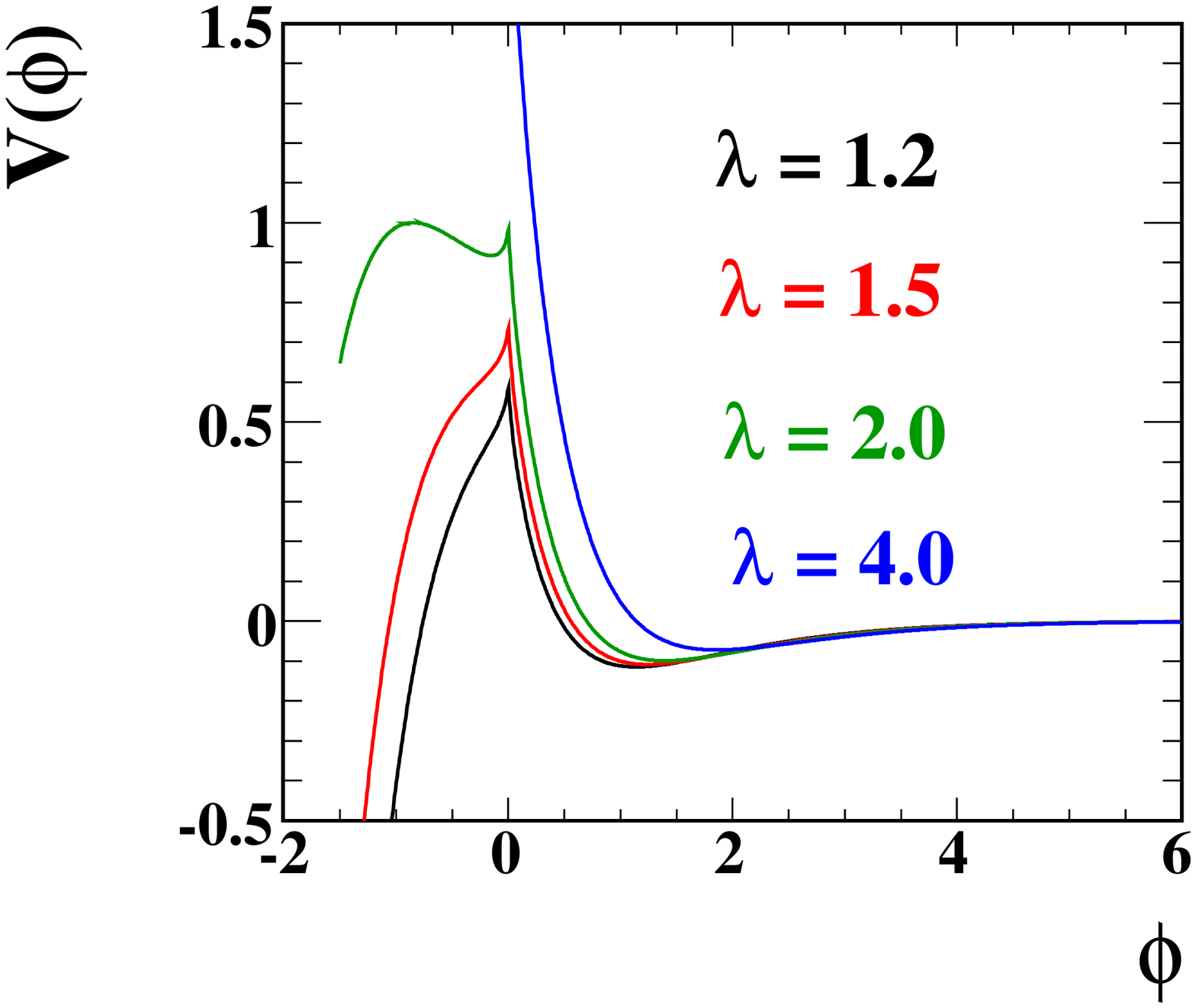}
\includegraphics[width = 5.5cm, height = 5cm]{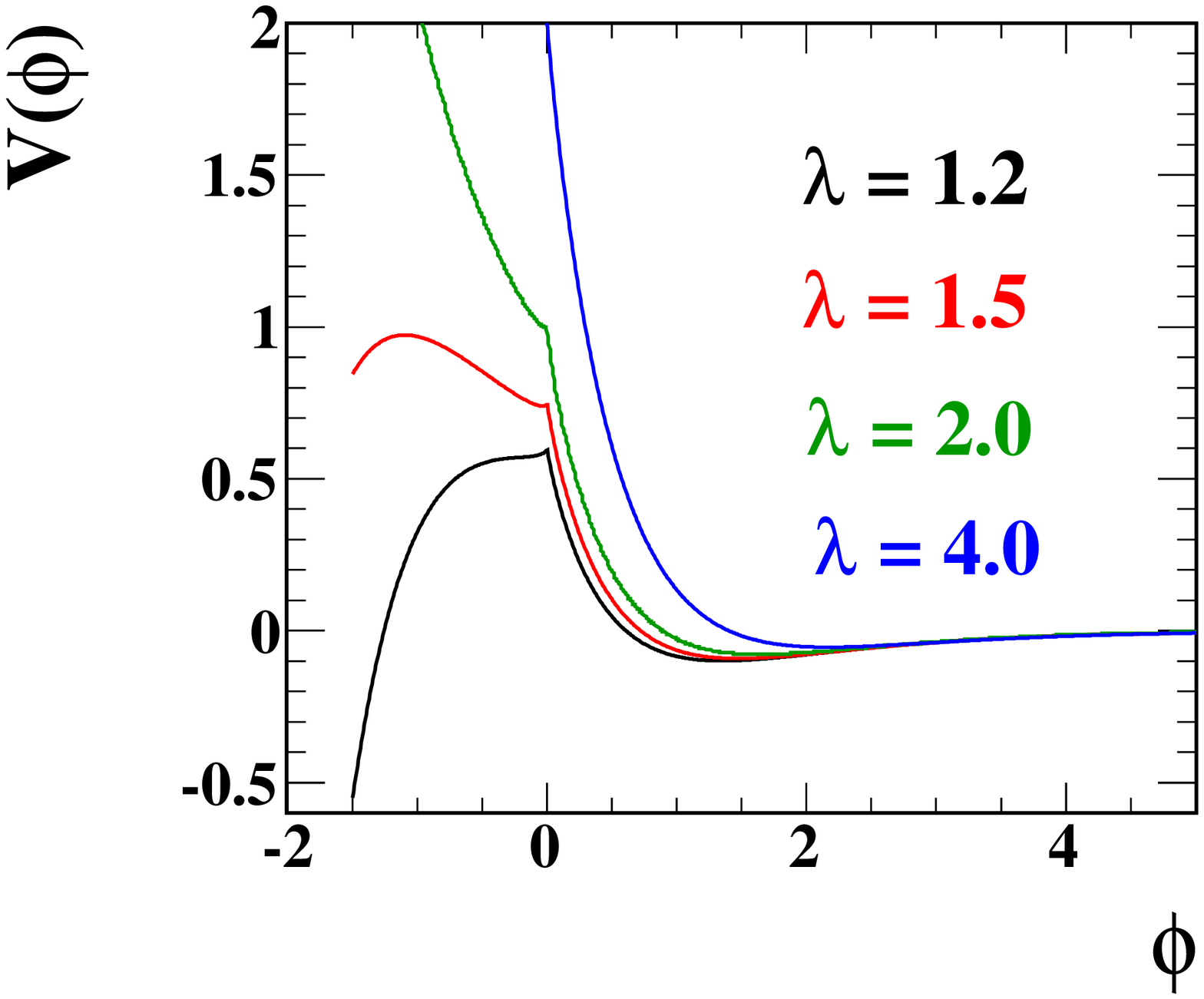}
\includegraphics[width = 5.5cm, height = 5cm]{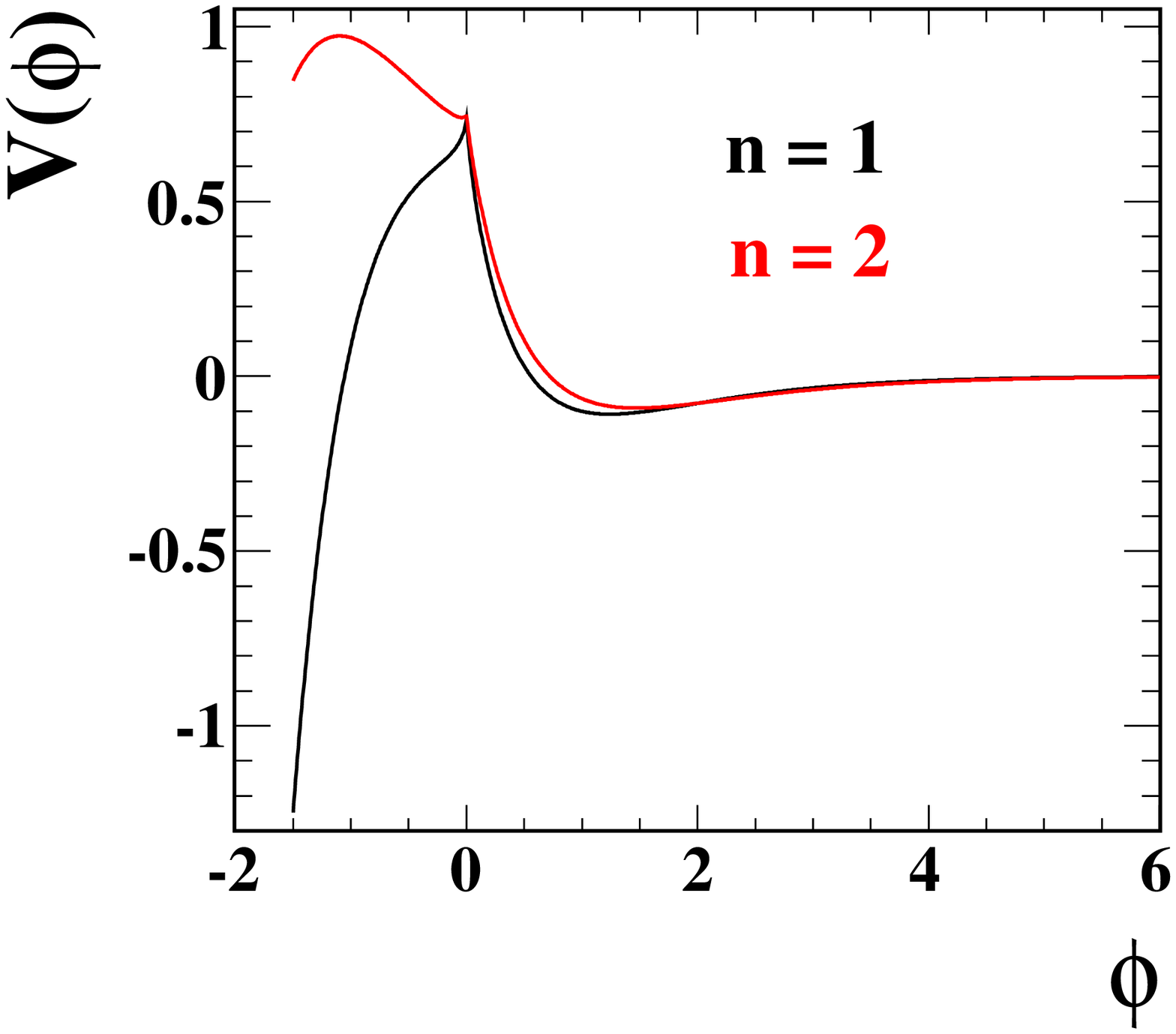}
}
\centerline{
\includegraphics[width = 5.5cm, height = 5cm]{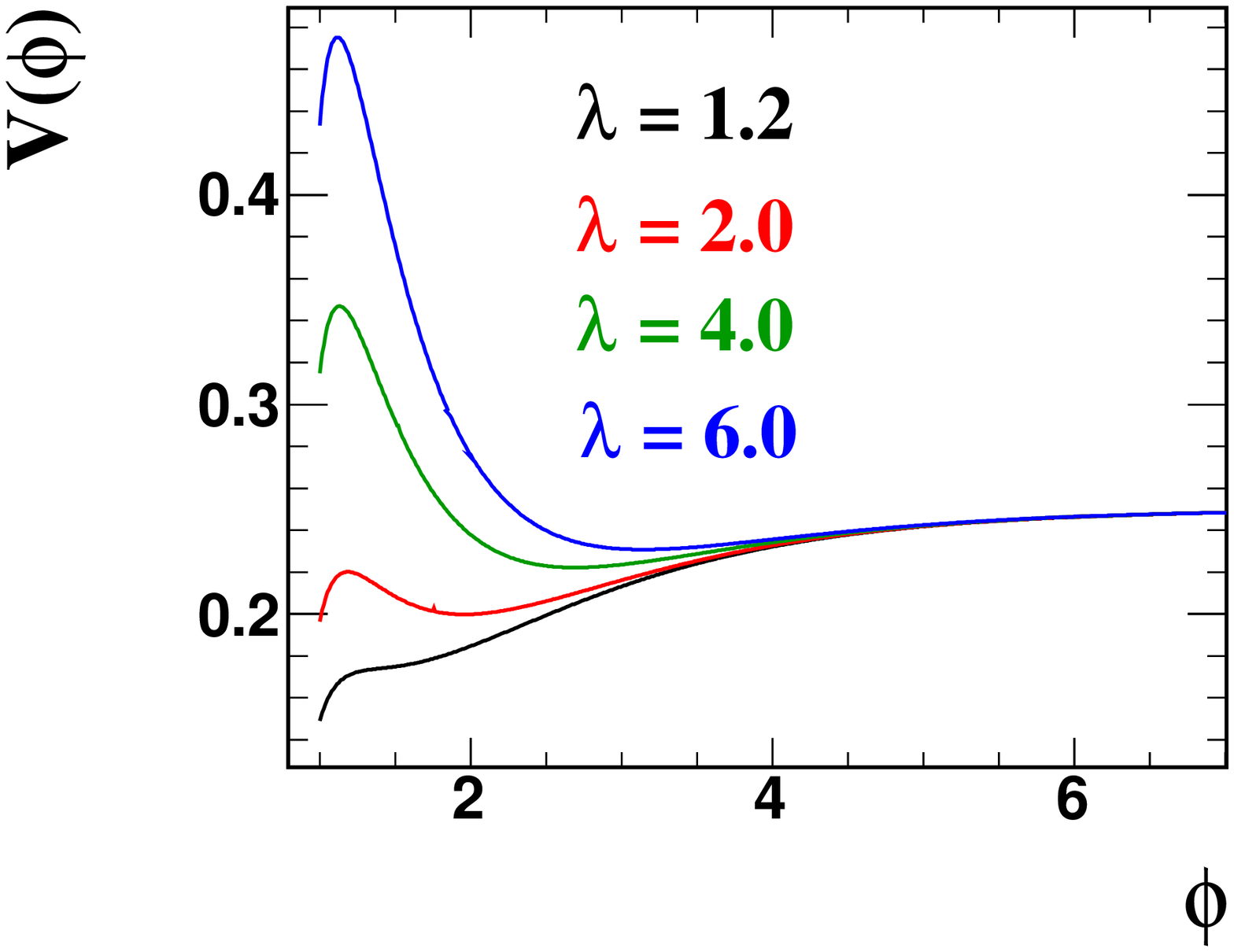}
\includegraphics[width = 5.5cm, height = 5cm]{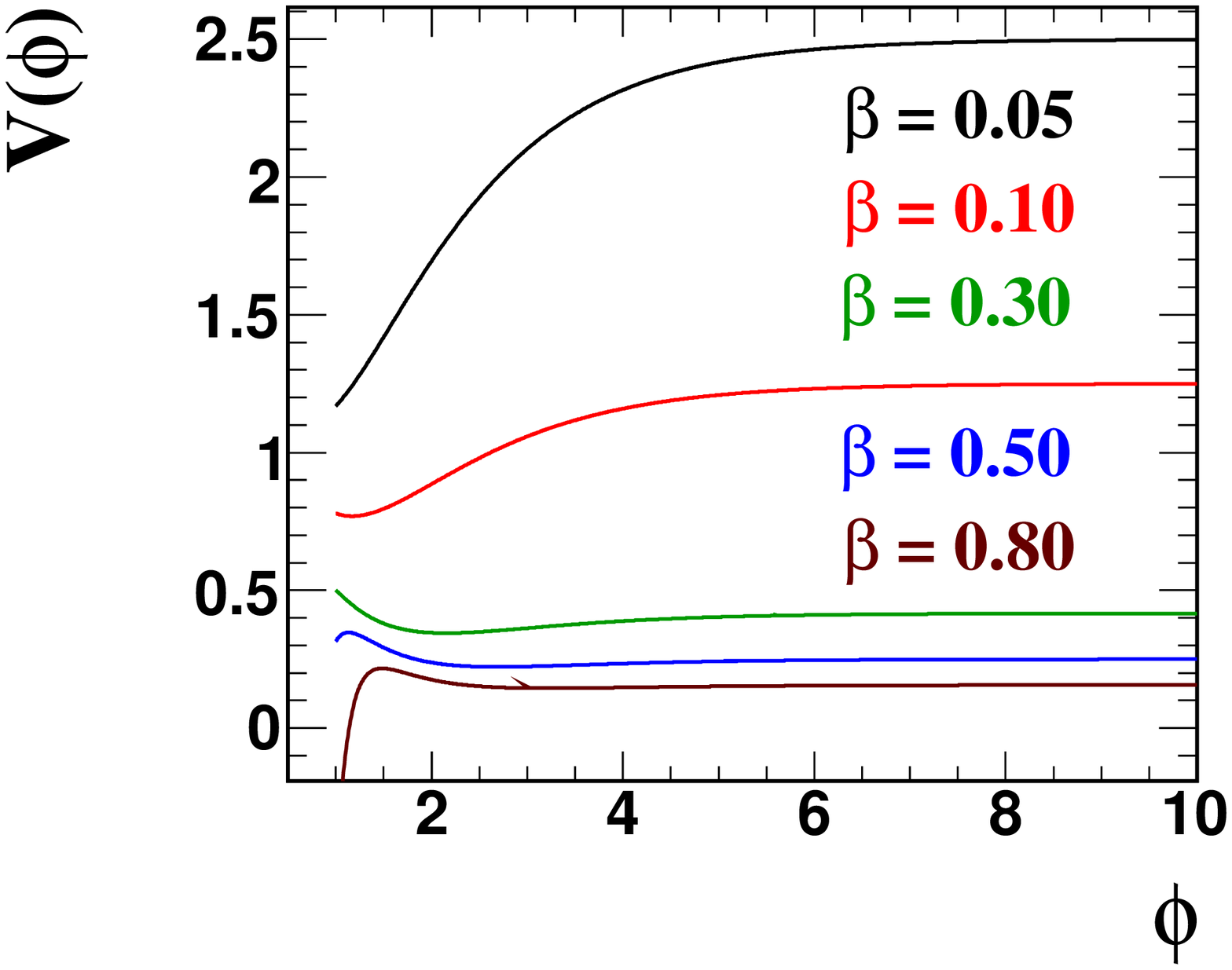}
}
\caption{Top panels: The behaviour of the field potential (\ref{eq28})
with respect to the field $\phi$ corresponding to the Starobinsky model 
(\ref{eq26}). The left panel is for $n = 1$ and the middle panel is for
$n = 2$ with $\lambda =  1.2$, $1.3$, $2.0$ and $4.0$. The right panel is 
for $n = 1$ and $2$ with $\lambda =  1.5$ to show exclusively the effect of
the parameter $n$ on the potential for a given value of $\lambda$. Bottom 
panels: The behaviour of the field potential (\ref{eq28a})
with respect to the field $\phi$ corresponding to the modified Starobinsky model
(\ref{eq26a}). The left panel is for $n = 2$ with $\beta = 0.5$ and the 
right panel is for $n = 2$ with $\lambda = 4$ to show exclusively the effect of 
the parameter $\beta$ on the potential for a given value of $\lambda$.}  
\label{fig5}
\end{figure}

The behaviours of the scalar field potential (\ref{eq28}) with respect to the 
field $\phi$ associated with the Starobinsky model (\ref{eq26}) for different
values of $n$ and $\lambda$ as mentioned above are shown in the top three 
panels of the Fig.\ref{fig5}. We observed that for smaller values of 
$\lambda$, when the 
field $\phi$ is rolling from negative to zero, the potential increases very 
fast from negative to its maximum positive value as the field reached its zero
value and then falls off rapidly to a almost flat region near its zero value 
when the field increases again from its zero value. This behaviour of the
potential for such $\lambda$ values form a pattern like the $\lambda$-pattern. 
Beyond this range of smaller values of $\lambda$, the $\lambda$-pattern of the 
potential vanishes gradually and the potential always remain 
positive for all values of the field $\phi$. The range of this smaller values 
of $\lambda$ depends on the value of $n$, higher the value of $n$, smaller is 
this range. Thus for higher values of $n$ and $\lambda$, the potential falls 
off from a very high value to its almost flat region near zero. Moreover, for 
all values of $n$ and $\lambda$ the potential becomes flat at a particular high 
positive value of $\phi$ depending on the value of $\lambda$, higher the value 
of $\lambda$, slightly higher this value of $\phi$. It should be noted that,
the unusual behavoiur of the field potential around the zero value of field 
$\phi$ for the lower and higher values of $\lambda$ depending on the value of
$n$ is the exhibition of the kind of singularity behaviour of the field $\phi$,
as mentioned above. 

The bottom two panels of the Fig.(\ref{fig5}) show as an example the behaviours 
of the field potential (\ref{eq28a}) with respect field $\phi$ associated with 
the modified Starobinsky model (\ref{eq26a}) for different values of $\lambda$ 
and $\beta$ with $n = 2$. It is clear from these panels that, there is no 
$\lambda$-pattern or very large value of the field potential for any value
of the parameter $\lambda$. From the bottom left panel, which is for $\beta = 
0.5$,
we see that for some lower value of the parameter $\lambda$,
the field potential increases slowly from its minimum to maximum value 
at a particular value of the field $\phi$ and then remains constant for all 
values of the field. Above such values of the parameter $\lambda$, the field
potential initially increases very fast to its highest peak value from a 
initial minimum and then falls to same minimum value at same value of the field
 $\phi$ as in the case of the lower value of the parameter $\lambda$ to behave
in a similar fashion for all parameter $\lambda$. The value of initial minimum
and the peak of the field potential increase with increasing value of the 
parameter $\lambda$. From the bottom right panel ($\lambda$ = 4.0) it is clear 
that, the initial behaviour of the field potential depends on the the value of 
the parameter $\beta$. For $\beta$ = 0.8, initial value of the field potential 
increases very fast from higher negative value to its positive peak value 
around the value of the $\phi$ = 1, which shows some unusual nature. However,
in maximum situations we may avoid the singularity behaviour of the scalar 
field potential by using the modified Starobinsky model (\ref{eq26a}). Moreover,
it should be noted that, scalar field $\phi$ associated with this model is
always positive as we have pointed above already. As already mentioned above,
still we are using the Starobinsky original model (\ref{eq26}) to continue
our discussions further, in view of our motivation of this work.               

\begin{figure}[hbt]
\centerline
\centerline{
\includegraphics[width = 7cm, height = 6cm]{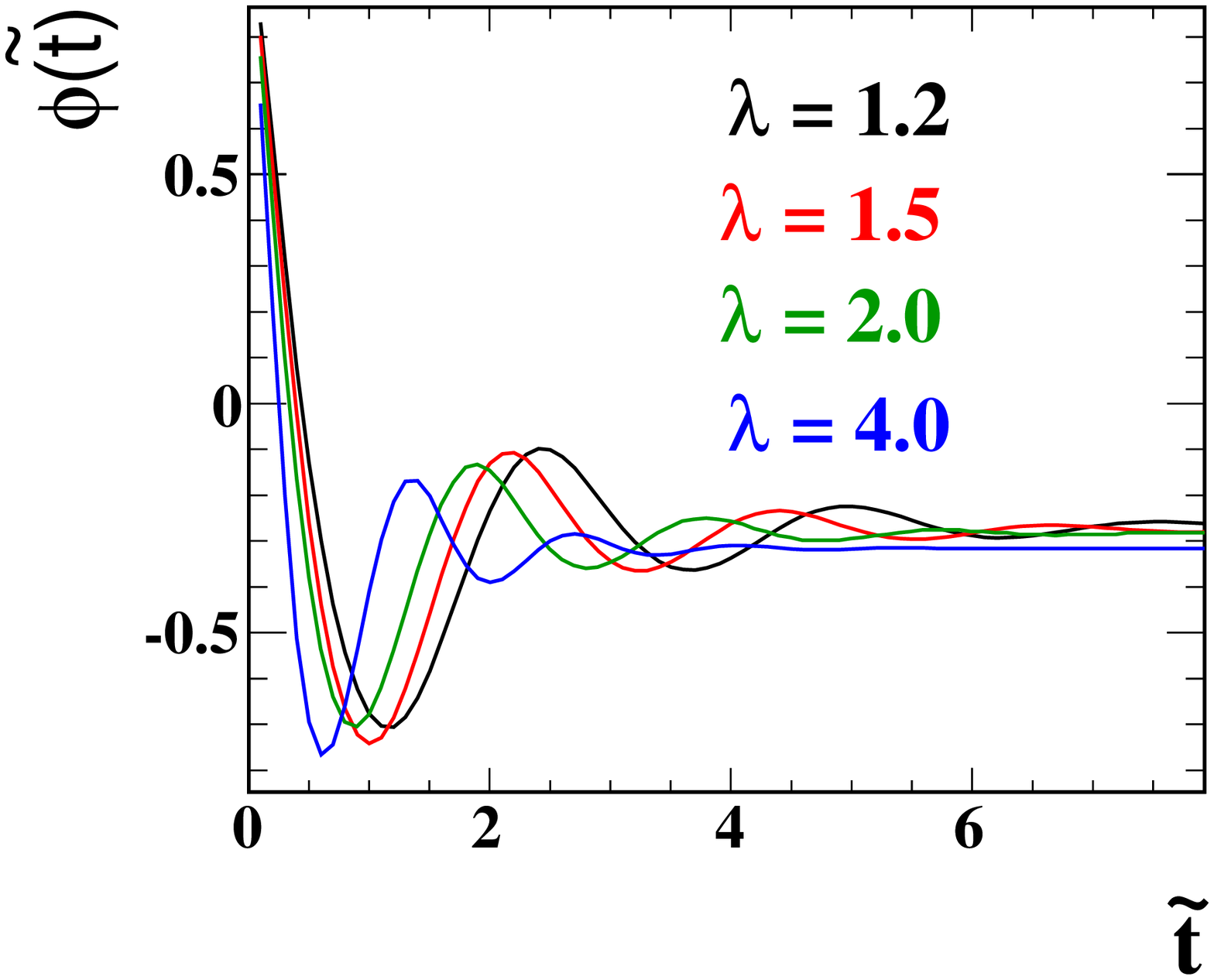}
\includegraphics[width = 7cm, height = 6cm]{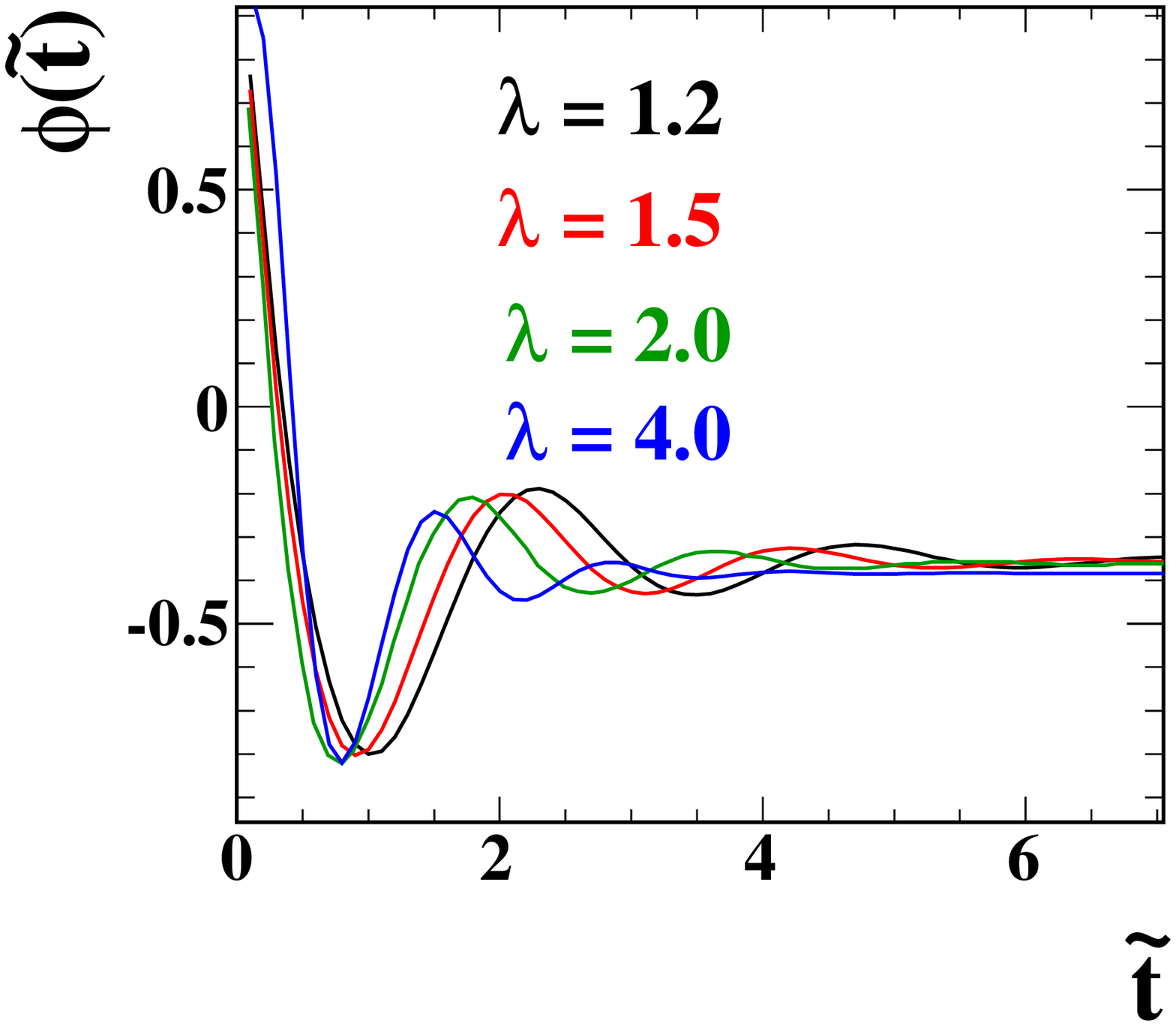}
\includegraphics[width = 7cm, height = 6cm]{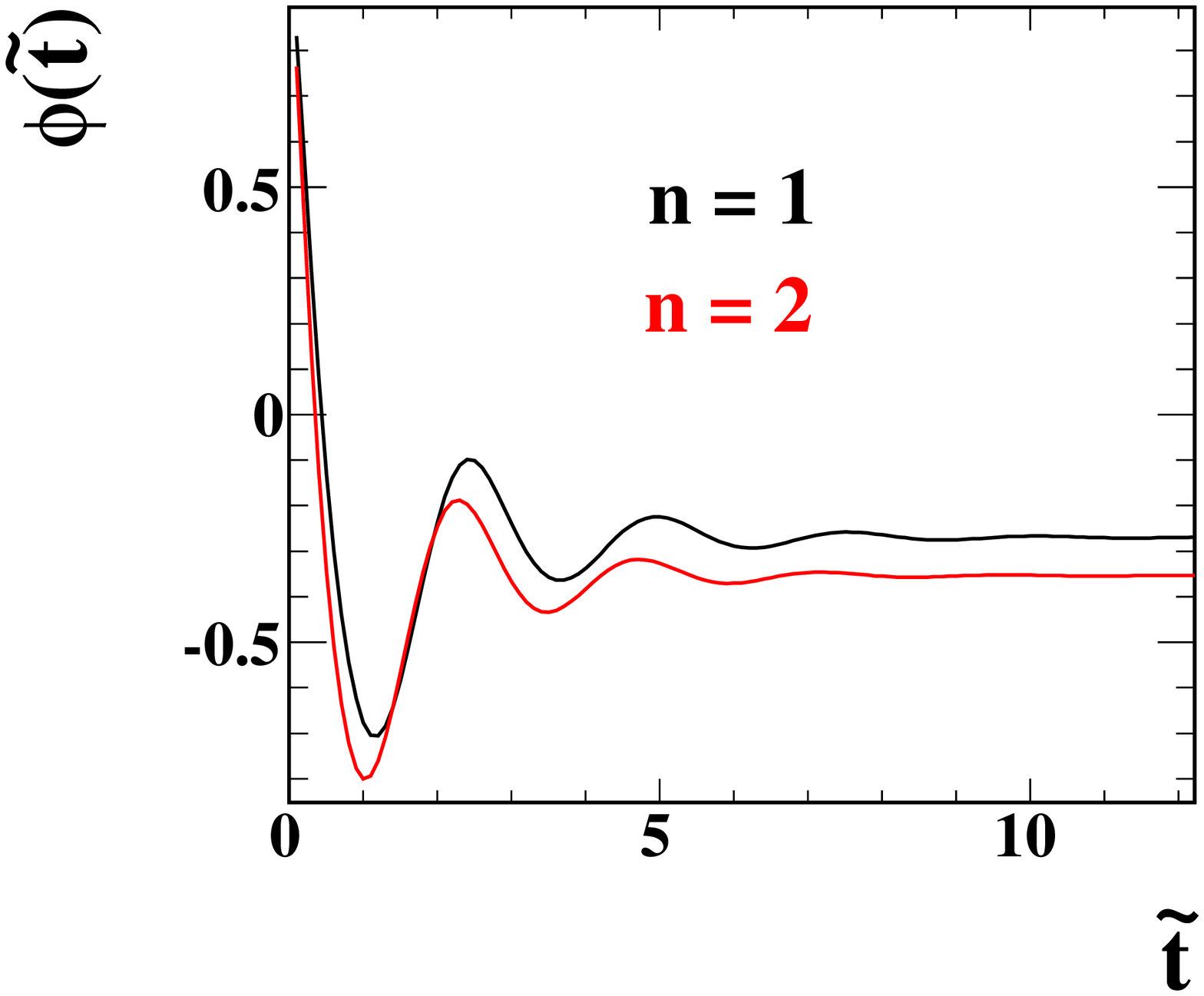}
\includegraphics[width = 7cm, height = 6cm]{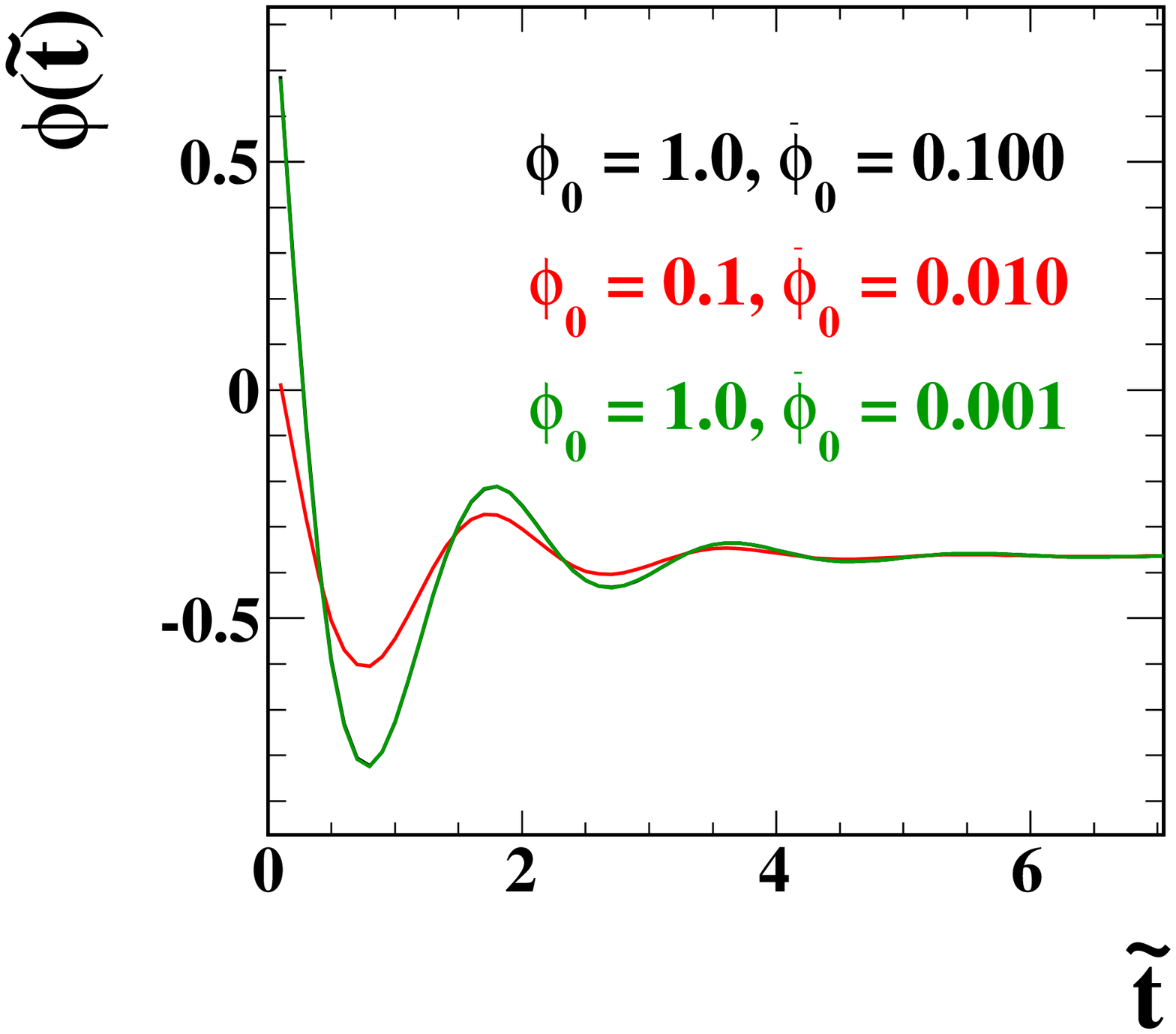}
}
\caption{Time evolution of the field $\phi$ obtained from numerical solutions 
of equations (\ref{eq16}) and (\ref{eq17}) using the field potential 
(\ref{eq28}) for different values of $n$ and $\lambda$. The top two panels are 
for $n=1$ (left) and $2$ (right) with different values of $\lambda$. The bottom
left panel is for $\lambda = 1.2$ with two different values of $n$. In all
these panels $\phi_0=1.0$ and $\dot{\phi}_0 = 0.1$ are used. While the bottom 
right panel is for different initial values of $\phi$ and $\dot{\phi}$ with 
$n = 2$ and $\lambda = 2.0$.}
\label{fig6}
\end{figure}

The Fig.\ref{fig6} shows the results of numerical solution of field equations
(\ref{eq16}) and (\ref{eq17}) using the field potential (\ref{eq28}) for 
different values of $n$ and $\lambda$. The initial values of field $\phi$ and
field velocity $\dot{\phi}$ are same as their values used in the case of 
power-law model. The general pattern of time evolution of the field is that the
field $\phi$ falls off rapidly from its initial positive value to its minimum
negative value and then oscillates for sometimes with a decreasing amplitude 
before becomes steady with a negative value. For higher values of $\lambda$ and
$n$ the field is legging behind the field for the lower values of these 
parameters by shifting its minimum towards more negative side. This indicates
that the magnitude of the field is smaller for higher values of $\lambda$ and 
$n$ in comparison to the field of their smaller values. It should be noted that
for the given $\lambda$ or the given $n$ this behaviour of the field $\phi$ is
observed for different values of $n$ or $\lambda$. However, in the case of
$\lambda = 4.0$ for $n = 2$ a slight different behaviour of the field is 
observed as seen from the top right panel of this figure wherein the field
falls from higher initial values in contrast with the behaviour for
other cases. The initial field velocity has noticeable effect on the 
time evolution of the field as it is observed from the bottom right panel of 
this figure that for the lower value of the initial field 
velocity the field magnitude increases for a period before the field $\phi$
becomes steady with the same value attained for other initial conditions. 

\begin{figure}[hbt]
\centerline
\centerline{
\includegraphics[width = 5.5cm, height = 5cm]{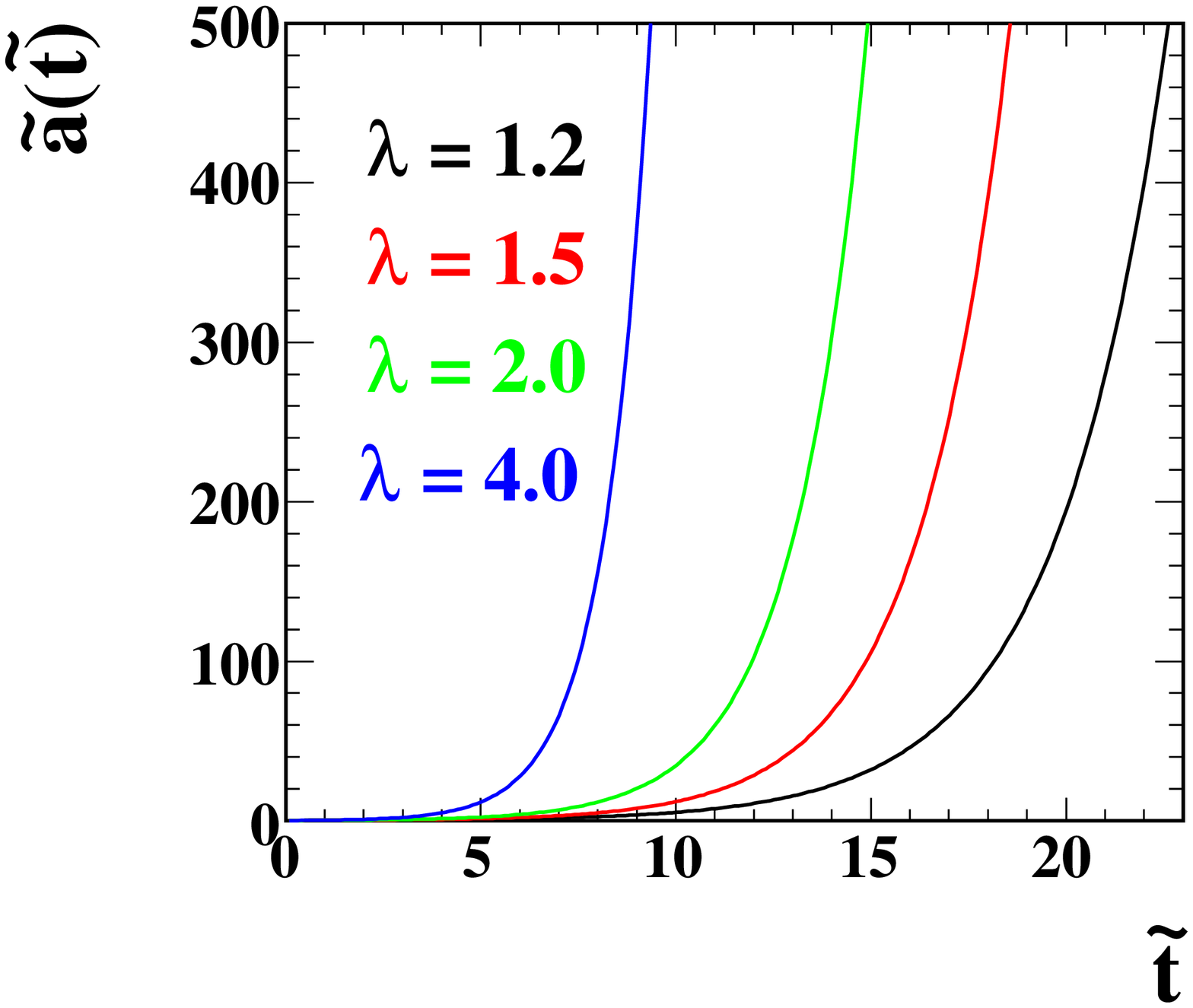}
\includegraphics[width = 5.5cm, height = 5cm]{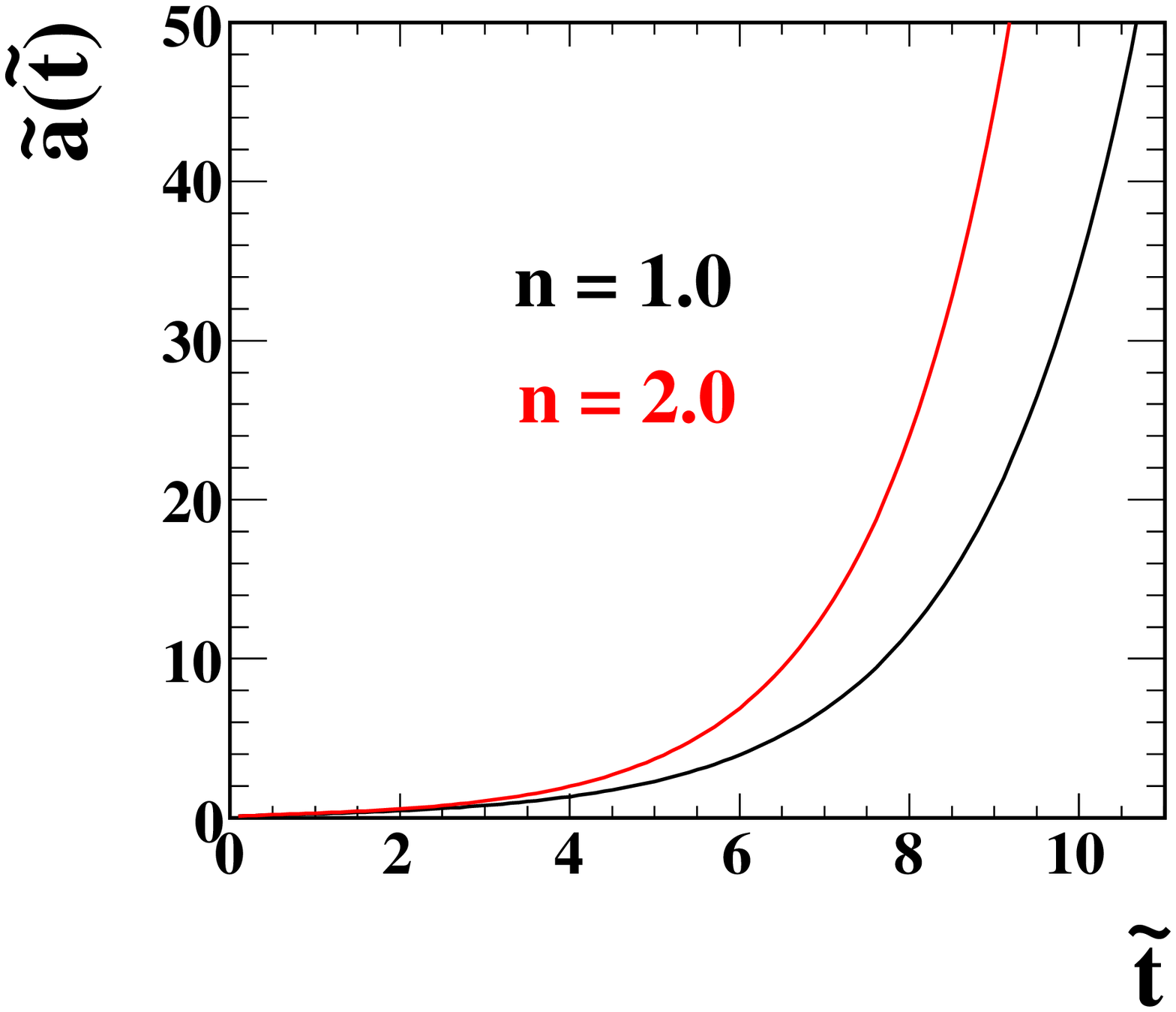}
\includegraphics[width = 5.5cm, height = 5cm]{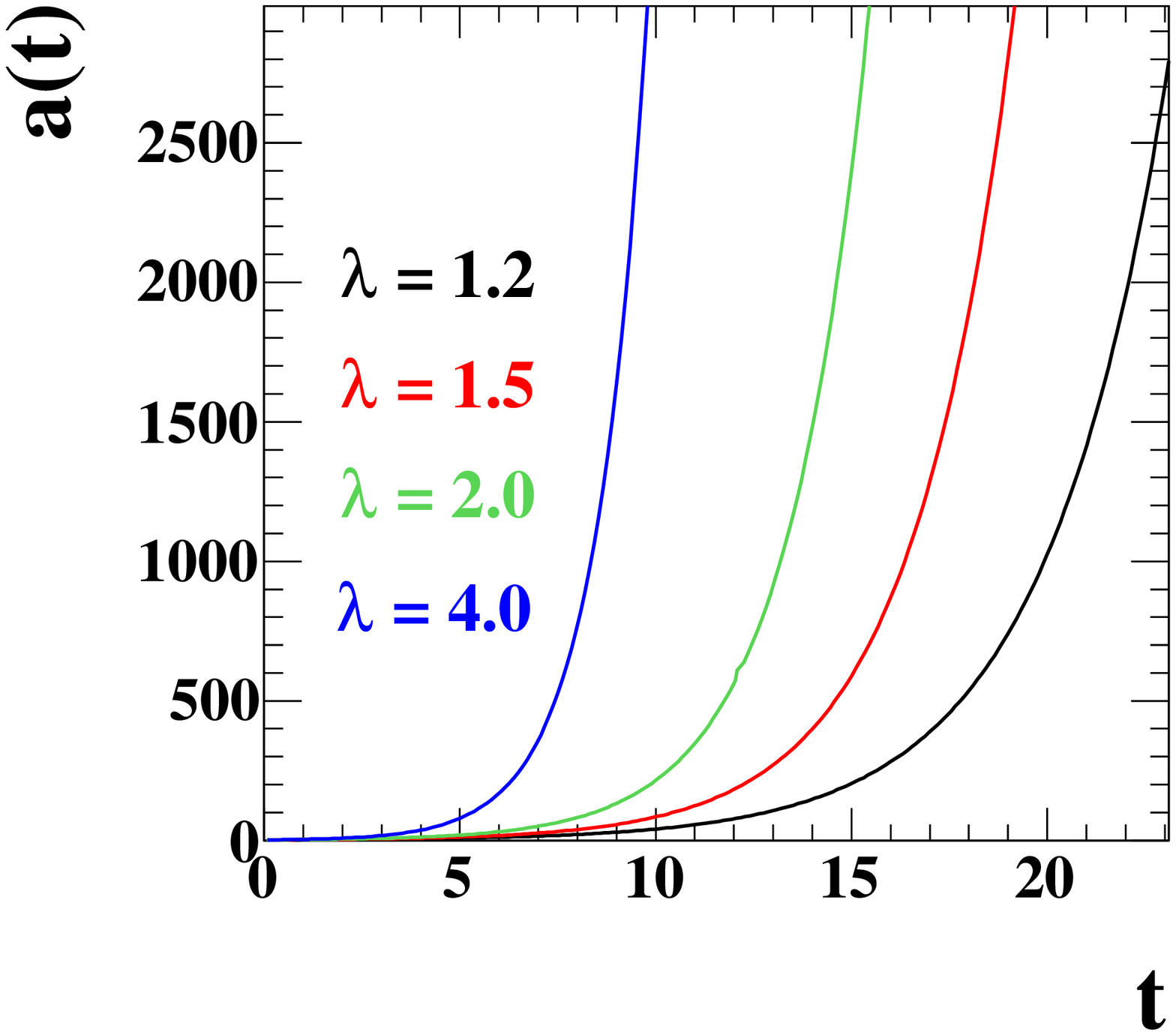}
}
\caption{First two panels: Evolution of the scale factor in the Einstein 
frame obtained from the numerical solutions of equations (\ref{eq16}) and 
(\ref{eq17}) using the field potential (\ref{eq28}) for different values of 
$n$ and $\lambda$. The first panel is for $n=1$ with different values of 
$\lambda$. Whereas, the second panel is for $\lambda = 2.0$ with two vales of 
$n$. In both these panel $\phi_0=1.0$ and $\dot{\phi}_0 = 0.1$ are used. 
Last panel: Evolution of the scale factor in the Jordan frame obtained from the
first panel by using the equation (\ref{eq20}).}
\label{fig7}
\end{figure}

The study of the scale factor evolution in the Einstein frame using the 
scalar degree of freedom of this model, as we have studied in the case of the 
power-law $f(R)$ model, is shown in the first two panels of the Fig.\ref{fig7}. 
It is clear from the fist panel of this figure that, during a considerable 
initial period the expansion of the Universe is slower and after that period 
the Universe experiences very fast accelerated expansion similar to the 
exponential expansion under this model. This initial period becomes shorter as 
the value of $\lambda$ increases. Moreover, for a given value of $\lambda$ 
this slower expansion period of the Universe is shorter for the higher value 
of $n$ as shown in the second panel of this figure. That is higher the value 
of $n$ and $\lambda$, the earlier is the period of accelerated expansion of the 
Universe. Thus this model could explain the late time accelerated expansion of 
the Universe with the suitable values of $n$ and $\lambda$.

For the same purpose as in the case of power law $f(R)$ gravity model, here 
also we have calculated the cosmic time $t$ and the scale factor $a(t)$ in 
the Jordan frame for all conditions of the first panel of the Fig.\ref{fig7}. 
The results of the calculation are shown in the last panel of the the same 
figure. It shows that, the Starobinsky $f(R)$ gravity model (\ref{eq26}) 
produces the accelerated expansion of the Universe in the Jordan frame also 
with exactly similar in patterns for the values of the parameter 
$\lambda$ as in the case of the Einstein frame, but with a many times higher
in magnitude, which increases with increasing cosmic time.    

\begin{figure}[hbt]
\centerline
\centerline{
\includegraphics[width = 5.5cm, height = 5cm]{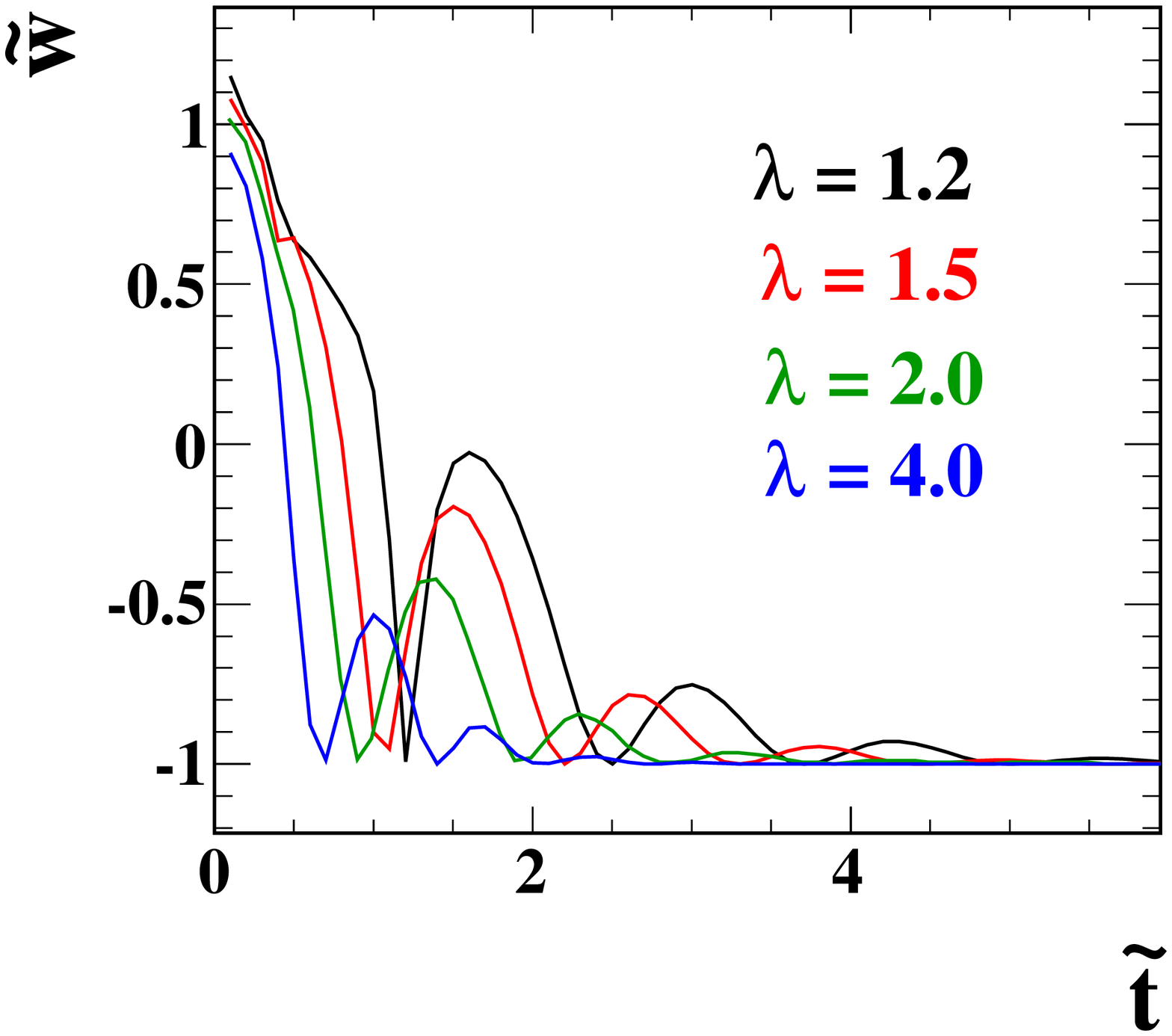}
\includegraphics[width = 5.5cm, height = 5cm]{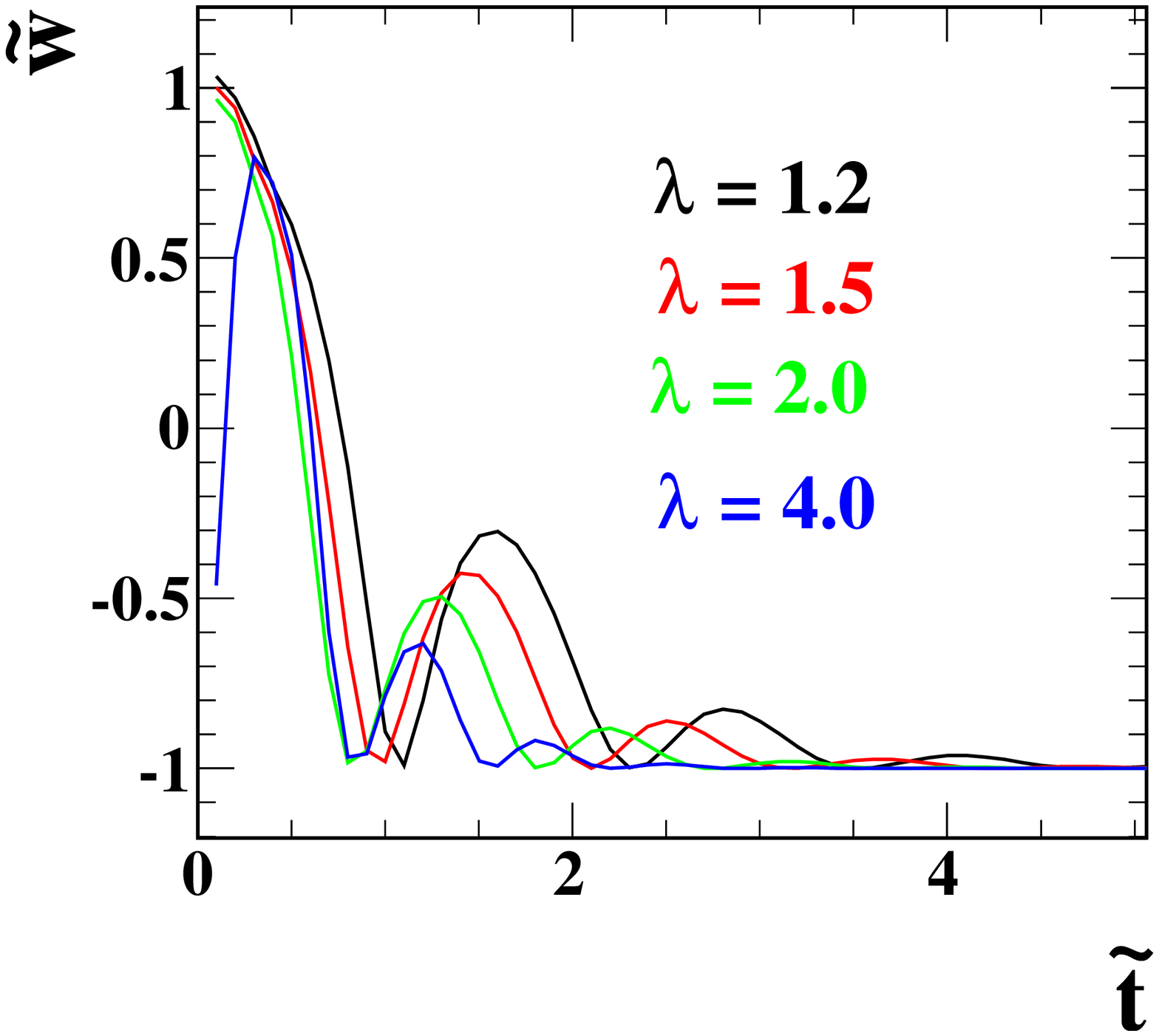}
\includegraphics[width = 5.5cm, height = 5cm]{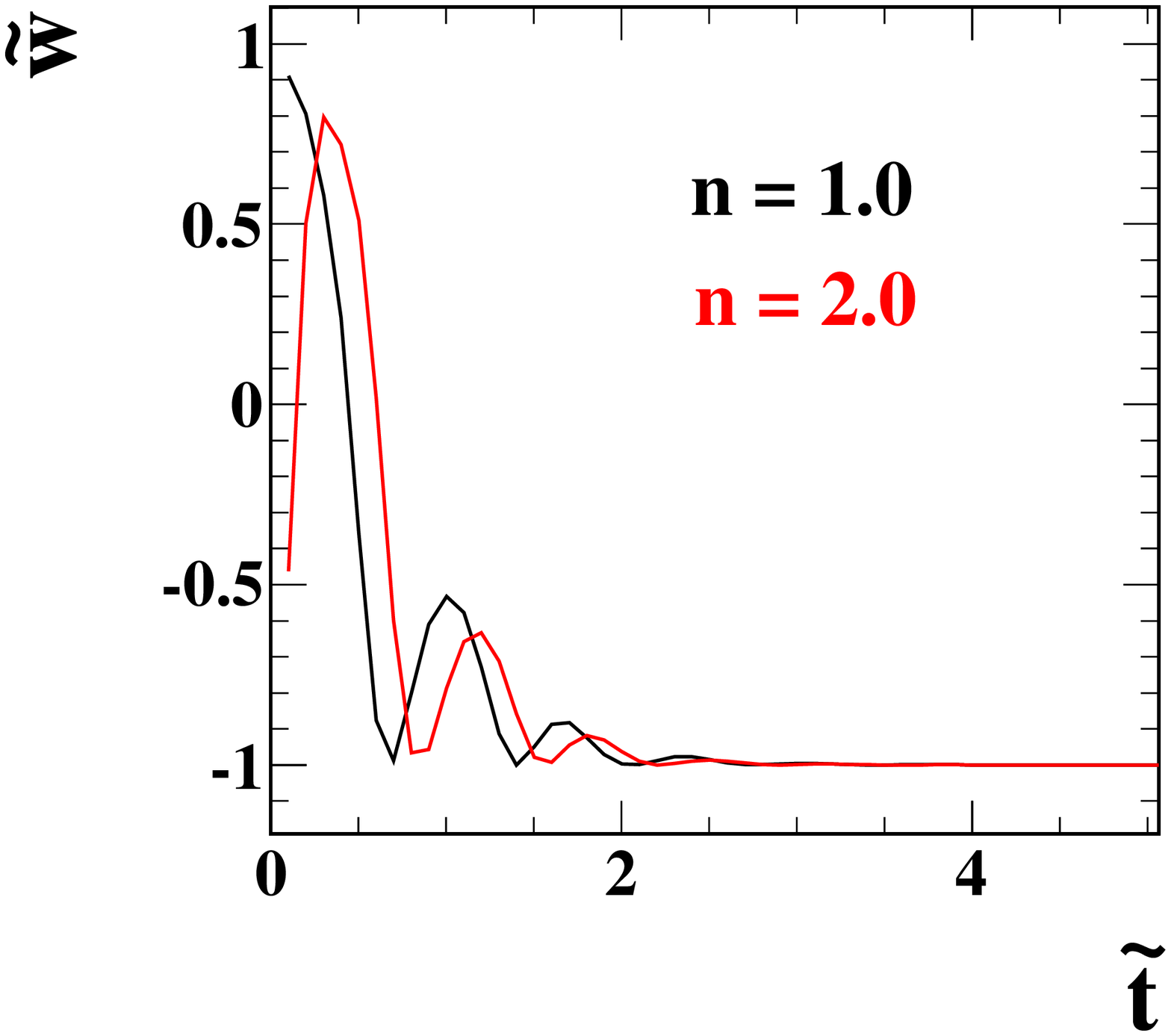}
}
\caption{Equation of state of the scalar field in the Einstein frame obtained 
from the numerical solutions of equations (\ref{eq16}) and (\ref{eq17}) using
the field potential (\ref{eq28}) for different values of $n$ and $\lambda$. The
first two panel is for $n=1$ and $2$ respectively with different values of
$\lambda$. Whereas the last panel is for $\lambda = 4.0$ with two vales of $n$. All initial conditions are same as in the previous figure.}
\label{fig8}
\end{figure}

As described in the previous Sec.III(A) for the power-law $f(R)$ gravity model, 
we have calculated in the same way the equation of state of the scalar field 
related with the Starobinsky $f(R)$ gravity model (\ref{eq26}). The results of
this numerical calculation is shown in the Fig.\ref{fig8}. The patterns of
time evolution of the equation state of the scalar field for 
different values of $\lambda$ and $n$ are almost similar to the patterns of 
time evolution of the field $\phi$ for theses parameters. That is the equation 
of state is also oscillate with decreasing amplitude during a period after 
falling from its initial positive value and then after that period it remains
steady with time. During a initial period depending on the value of 
$\lambda$ and $n$ the equation of state falls from $1$ to $-1$, which indicates
that during this period the scalar field corresponding to the model 
(\ref{eq26}) make a transition from free scalar field ($w = 1$) through matter
($w\approx 0$, radiation ($w = 1/3$) and dark energy ($w < -1/3$) to 
cosmological constant $\Lambda$ ($w = -1$). During the oscillation period 
the scalar field behaves as matter to the cosmological constant and finally
behaves as the cosmological constant $\Lambda$ after completing this period
depending on the values of the parameters $\lambda$ and $n$. There is a special
behaviour of the equation of state for the $n=2$ and $\lambda = 4.0$ during 
the initial
period. In contrast to the behaviour of the equation of states for the other
pairs of parameters, the equation of state of this pair of parameters initially
make transit from the dark energy state to other states and then make
transition to the cosmological constant state. However for latter periods
the behaviour of equation of state for this pair of parameters is similar
to the other cases.

\section{Conclusion}
We have investigated the scalar degrees of freedom and it's related 
cosmological implications of two $f(R)$ gravity models, viz., (i) the power-law
$f(R)$ gravity model $\xi R^n$ and (ii) the Starobinsky disappearing 
cosmological constant $f(R)$ gravity model for different values of the 
parameters of these models. This study is basically on the theoretical 
interests rather than the observational consequences. The power-law $f(R)$ 
gravity model is considered due to the existence it's power-law solutions and 
consequently absence of singularity problem \cite{Goheer}. On the other hand 
the Starobinsky $f(R)$ gravity model with the disappearing cosmological 
constant \cite{Starobinsky} is a standard $f(R)$ gravity model which provide 
viable cosmology by satisfying the solar system and the laboratory tests. As 
the scalar degree of freedom of $f(R)$ gravity model can be treated explicitly 
in the Einstein frame, so we used this frame for this study by conformal 
transformations from the Jordan frame.

It is found that the field potential of the power-law $f(R)$ gravity model is 
the well behaved function of the the field $\phi$ representing the scalar 
degree of freedom of the model and increases very fast as $\phi$ increases, 
which depends on the model parameters 
$\xi$ and $n$ such that for the lower value of them such process is faster than
higher value. From the numerical simulation of the field equations, we found
that $\phi$ falls towards the negative values as time increases, which is more 
faster during the initial period of time than late times. The rate of falling
the value of the field slower for the higher values of the parameter $\xi$.
However $n$ does not have much impact as $\xi$ has on the behaviour of field
$\phi$ with respect to time. Initial conditions on the field and field velocity
have negligible effect on the nature of the time evolution of $\phi$.

In the case of the Starobinsky $f(R)$ gravity model with disappearing 
cosmological constant the field potential shows a peculiar behaviour with 
respect to the field $\phi$. When the field increases from its negative value 
to zero, the field potential increases from its lowest negative value to the 
maximum positive value and then falls off very fast to the flat region near to 
its zero value as the field increases from zero to higher positive values for
a lower value range of the parameter $\lambda$. This range of the lower value
of the parameter $\lambda$ decreases with increasing the value of $n$. This 
behaviour of the filed potential form a pattern similar to the 
$\lambda$-pattern. The $\lambda$-pattern gradually vanishes with increasing 
values of the parameters $\lambda$ and $n$. For higher values of the field 
$\phi$ the values of field potential for different $\lambda$ and $n$ parameters
are almost identical with a slight dependence on the value of the parameter 
$\lambda$. All unusual behaviours of the field potential can be avoided
by adding a term $\propto R^2$ to the Starobinsky model. 
The numerical solution for the time evolution of the field $\phi$
for this model shows that the field falls off rapidly from its maximum positive
value to its minimum negative value in the earlier period and then starts 
oscillating with a decreasing amplitude with time for a period before become 
steady
with time taking a particular negative value. This pattern of time evolution 
of the field $\phi$ depends on the the model parameters in such a way that with 
increasing values of $\lambda$ and $n$ the field $\phi$ leg behind its values
for lower model parameters with decreasing amplitudes to attain steady value
much earlier.           

The study
of the time evolution of the scale factor in the Einstein frame shows that
the Universe is accelerating very fast for both of our $f(R)$ gravity models,
which depends on values of the parameters of the models. For the case of the
power-law $f(R)$ model the scale factor depends on the parameters $\xi$ and 
$n$ in such a way that accelerating process slows down for higher values of 
$\xi$ and for lower values of $n$. Whereas in the case of the Starobinsky
$f(R)$ model the accelerating process becomes more faster for the higher value
of both parameters $\lambda$ and $n$. Moreover, in this case there is a initial
period depending on the value of $\lambda$ in which the expansion of the 
Universe is very slow. This initial period is shorter for the higher value of 
the parameter $\lambda$ and does not depends on the parameter $n$. After this
initial period the Universe is expanding with a very fast acceleration process.
This indicates that this model with the suitable model parameters $\lambda$ and 
$n$ is very effective to give the late time accelerated expansion of the 
Universe. We found that, the Universe expands almost in similar fashions in
the Jordan frame also as in the case of the Einstein frame for both these 
models, but with slightly higher in magnitude for the power law model and with
much higher in magnitude for the Starobinsky model.     

The equation of state of the scalar field for the power-law model is always 
negative and less than $-1/3$, which corresponds to the behaviour of the dark 
energy that produces the accelerated expansion of the Universe. The equation 
of state of the field for this model tends to $-1$ as the value of $n$ 
increases towards higher values, nearly equal to $2$. Thus for such values of 
$n$, the scalar field for the model behave as the cosmological constant 
$\Lambda$. The same study for the scalar field of the Starobinsky model shows
that the equation of state oscillates with a decreasing amplitude with time 
for a certain period after falling from its maximum value around $1$ to
its minimum value at $-1$. After this period the equation of state takes the
value equal to $-1$. This implies that the scalar field related with the 
Starobinsky disappearing cosmological constant $f(R)$ gravity model passages
through different phases before finally behaves as the cosmological constant 
$\Lambda$. The behavior of the time evolution and the dependence of the 
equation of state on the model parameters $\lambda$ and $n$ are almost similar
to the field $\phi$ itself.

At last it is interesting to mention that, our study can be applied to more
complicated, unified (inflation plus cosmic acceleration) $f(R)$ models of
the form given in \cite{Cognola}, to see more explicitly the stages of the
inflation and the cosmic acceleration of our Universe by using the scalar 
degree of freedom in Einstein frame. From the observational point of view, 
the results can be easily transformed to Jordan frame (which
is clear from our discussion in the previous sections), as by convention the 
Jordan frame is considered as physical frame, although it is still 
pre-mature to say, which frame is the realistic one, as far as the present 
cosmological observations are concerned. We are planing to work on these 
models and issues in near future.

\end{document}